\documentclass[journal,onecolumn]{IEEEtran}

\usepackage{amssymb,amsmath,amsfonts,psfrag,graphicx,setspace,subfigure,cite}

\def\eps{{ \epsilon }}
\newtheorem{theorem}{Theorem}[section]
\newtheorem{conjecture}[theorem]{Conjecture}
\newtheorem{lemma}{Lemma}[section]
\newtheorem{corollary}[lemma]{Corollary}
\newtheorem{definition}[theorem]{Definition}

\newtheorem{remark}[theorem]{Remark}
\newtheorem{comment}[theorem]{Comment}
\def\bC{{ \mathbf{C} }}
\def\bD{{ \mathbf{D} }}
\def\B{{ \mathcal{B} }}
\def\C {{ \mathcal{C} }}
\def\Co {{ \mathcal{C}o }}
\def\Cu{{ \mathcal{C}_u }}
\def\Cx{{ \mathcal{C}_x }}
\def\Cv{{ \mathcal{C}_v }}

\def\D{{ \mathcal{D} }}
\def\E{{ \mathcal{E} }}
\def\L{{ \mathcal{L} }}
\def\Mx{{ \mathcal{M}_x }}
\def\My{{ \mathcal{M}_y }}
\def\Mc{{ \mathcal{M}_c }}

\def\P {{ \mathcal{P} }}

\def\Pi {{ \mathcal{P}_{in} }}
\def\Po {{ \mathcal{P}_{out} }}
\def\Pm {{ \mathcal{P}_{mix} }}
\def\Q {{ \mathcal{Q} }}
\def\R{{ \mathcal{R} }}

\def\Ri {{ \mathcal{R}_{in} }}
\def\Ro {{ \mathcal{R}_{out} }}
\def\Rm {{ \mathcal{R}_{mix} }}
\def\bR{{ \mathbf{R} }}

\def\T{{ \mathcal{T} }}
\def\U {{ \mathcal{U} }}
\def\V{{ \mathcal{V} }}

\def\X{{ \mathcal{X} }}
\def\Y{{ \mathcal{Y} }}
\def\bU{{ \mathbf{U} }}
\def\bV{{ \mathbf{V}  }}
\def\bX{{ \mathbf{X}  }}
\def\bY{{ \mathbf{Y} }}

\def\ba {{ \mathbf{a} }}
\def\bb {{ \mathbf{b} }}
\def\bc {{ \mathbf{c} }}

\def\mh{{ \hat{m} }}
\def\u {{ \mathbf{u} }}
\def\v {{ \mathbf{v} }}
\def\wh {{ \hat{w} }}
\def\x {{ \mathbf{x} }}
\def\y {{ \mathbf{y} }}

\def\a {{ \alpha }}
\def\b {{ \beta }}
\def\g {{ \gamma }}
\def\d {{ \delta }}
\def\t {{ \theta }}
\def\eps {{ \epsilon }}

\def\ve {{ \varepsilon }}

\def\bx {{ \mathbf{x} }}
\def\by {{ \mathbf{y} }}

\def\calD {{ \cal{D} }}

\def\ri {{ r_{in} }}
\def\ro {{ r_{out} }}

\def\ra{{ \rightarrow }}

\def\qb {{ \bar{q} }}
\def\tb {{ \bar{\theta} }}
\def\rxy {{ \rho_{xy} }}
\def\rxu {{ \rho_{xu} }}
\def\rux {{ \rho_{ux} }}
\def\ryv {{ \rho_{yv} }}
\def\rxv {{ \rho_{xv} }}
\def\ryu{{ \rho_{yu} }}
\def\ruv {{ \rho_{uv} }}

\def\calH {{ \cal{H} }}

\def\boldC{{\mathbf{C}}}

\def\boldX{{\mathbf{X}}}
\def\boldY{{\mathbf{Y}}}

\begin{document}
\title{Achievable rates for pattern recognition}
\author{M.~Brandon~Westover~
        and~Joseph~A.~O'Sullivan,~\IEEEmembership{Fellow,~IEEE}%
\thanks{This work was supported by the Mathers
Foundation and by the Office of Naval Research.}%
} \markboth{}{Westover and O'Sullivan: Achievable rates for pattern recognition.}
%\tableofcontents
%\newpage
\maketitle

% Beware--changing this will change the appearance of some of the figures!
\setlength{\unitlength}{1.5cm}

\begin{abstract}
Biological and machine pattern recognition systems face a common
challenge: Given sensory data about an unknown object, classify
the object by comparing the sensory data with a library of
internal representations stored in memory. In many cases of
interest, the number of patterns to be discriminated and the
richness of the raw data force recognition systems to internally
represent memory and sensory information in a compressed format.
However, these representations must preserve enough information to
accommodate the variability and complexity of the environment, or
else recognition will be unreliable. Thus, there is an intrinsic
tradeoff between the amount of resources devoted to data
representation and the complexity of the environment in which a
recognition system may reliably operate.

In this paper we describe a general mathematical model for pattern
recognition systems subject to resource constraints, and show how
the aforementioned resource-complexity tradeoff can be
characterized in terms of three rates related to number of bits
available for representing memory and sensory data, and the number
of patterns populating a given statistical environment. We prove
single-letter information theoretic bounds governing the
achievable rates, and illustrate the theory by analyzing the
elementary cases where the pattern data is either binary or
Gaussian.
\end{abstract}

%\tableofcontents

\section{Introduction}
\PARstart{P}{attern} recognition is the problem of inferring the
nature of unknown objects from incoming and previously stored
data. In real-world operating environments, the volume of raw data
available often exceeds a recognition system's resources for data
storage and representation. Consequently, data stored in memory
only partially summarizes the properties of physical objects, and
internal representations of incoming sensory data are likewise
imperfect approximations. In other words, pattern recognition with
physical systems is frequently a problem of inference from
\emph{compressed data}. However, excessive data compression
precludes reliable pattern recognition. In this paper we attempt
to answer the following question: In a given environment, what are
the least amounts of memory data and sensory data consistent with
reliable pattern recognition?

The paper is organized as follows. In section \ref{sec: informal
description} we introduce the general problem qualitatively.
Relationships between the present work and other pattern
recognition research is briefly described in section \ref{sec:
related work}. In section \ref{sec:Formal Problem statement} we
formalize our problem as that of determining which combinations of
three key rates are achievable, that is, which rate combinations
are consistent with the possibility of reliable pattern
recognition. These rates are directly related to number of bits
available for representing memory and sensory data, and the number
of distinct patterns which the recognition system must be able to
discriminate. The main results of the paper are single letter
formulas providing inner and outer bounds on the set of achievable
rates, given in section \ref{Sec: main results} and discussed in
section \ref{Sec: discussion}. The theory is illustrated by
applying it to the Binary case in section \ref{sec: Binary case}
and the Gaussian case in \ref{sec: Gaussian case}.

\section{Informal problem description}\label{sec: informal description}
\begin{figure}
\centering
\begin{picture}(5.5,5.5)(0,0)

\put(0,4.5){\framebox(1,1){ $p(\bx)$ }}
\put(0.1,4.5){\vector(0,-1){.5}} \put(0.2,4.5){\vector(0,-1){.5}}
\put(0.4,4.25){\makebox(0,0)[l]{$\hdots$}}
\put(0.9,4.5){\vector(0,-1){.5}} \put(0,3){\framebox(1,1){$
\mathcal{C}_x $}} \put(0.1,3){\vector(0,-1){.5}}
\put(0.2,3){\vector(0,-1){.5}}
\put(0.4,2.75){\makebox(0,0)[l]{$\hdots$}}
\put(0.9,3){\vector(0,-1){.5}} \put(0,1.5){\framebox(1,1){$
f_n(\cdot) $}} \put(0.1,1.5){\vector(0,-1){.5}}
\put(0.2,1.5){\vector(0,-1){.5}}
\put(0.4,1.25){\makebox(0,0)[l]{$\hdots$}}
\put(0.9,1.5){\vector(0,-1){.5}} \put(0,0){\framebox(1,1){
$\mathcal{C}_u$ }} \put(1,3.8){\vector(1,0){1}}
\put(1,3.7){\vector(1,0){1}}
\put(1.5,3.2){\makebox(0,0)[b]{$\vdots$}}
\put(1,3.1){\vector(1,0){1}} \put(2,3){\framebox(1,1){
\shortstack{ Select \\ one } }} \put(2,4.5){\framebox(1,1){ $p(w)$
}} \put(2.5,4.5){\vector(0,-1){.5}}
\put(2.7,4.25){\makebox(0,0)[l]{$ W $}}
\put(3,3.5){\vector(1,0){1.5}} \put(3.75,3.7){\makebox(0,0)[b]{$
\bX(W) $}} \put(4.5,3){\framebox(1,1){ $p(\by|\bx)$ }}
\put(5,3){\vector(0,-1){.5}}
\put(5.2,2.75){\makebox(0,0)[l]{$\bY$}}
\put(4.5,1.5){\framebox(1,1){ $\phi_n(\cdot)$ }}
\put(4.5,2){\line(-1,0){2}} \put(3.5,2.5){\makebox(0,0)[t]{$\mu$}}
\put(2.5,2){\vector(0,-1){1}} \put(2,0){\framebox(1,1){
$g(\cdot,\cdot)$ }} \put(3,0.5){\vector(1,0){1}}
\put(3.5,0.7){\makebox(0,0)[b]{$\hat{W}$}}
\put(1,0.8){\vector(1,0){1}} \put(1,0.7){\vector(1,0){1}}
\put(1.5,0.2){\makebox(0,0)[b]{$\vdots$}}
\put(1,0.1){\vector(1,0){1}}
\end{picture}

\caption{Block diagram for a generic pattern recognition
system.}\label{Fig: MainDiagram}
\end{figure}
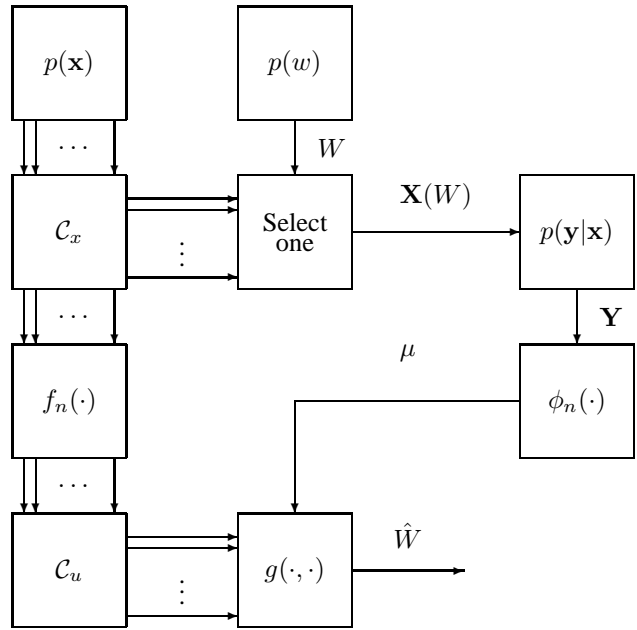

In general, statistical pattern recognition problems may be
specified in terms of a probabilistic model of the
\emph{environment} (`nature') \footnote{Non-probabilistic models
have also been considered. Arguments for preferring the
probabilistic formulation are discussed in
\cite{mumford99Dawning}.}; a pattern recognition \emph{system};
and the interactions of the system with the environment during two
distinct \emph{modes of operation}, a training (`offline') phase
and a testing (`online') phase. Informal descriptions for the
environment and system models we study are given below, and
formalized in section \ref{sec:Formal Problem statement}. Our
model and viewpoint are similar to others in the statistical
pattern recognition literature (see, e.g.
\cite{jain00statistical,ProbThPatRec,FukunagaBook,BishopNeuralNetBook,lecun-98}),
but fits most closely within the framework of Pattern Theory (see
e.g.
\cite{BayesPercIntro,Mumford96PABI,mumford94neuronal,KerstenMammasianYuille2004,grenanderElPatTheory}).
Please refer to the block diagram in figure \ref{Fig: MainDiagram}
while reading the following description.

\subsection{Environment}
\emph{Training patterns and the training phase.} The environment
for a pattern recognition system is defined as the set of distinct
entities that the system must learn to reliably distinguish. These
entities are hereafter referred to simply as \emph{patterns}, and
may include, for example, distinct physical objects, properties of
objects, or arrangements of multiple objects. We assume each
pattern can be represented by an $n-$vector
$\x=(x_1,x_2,\ldots,x_n)$ whose elements take values in some
alphabet $\X$. Of the $|\X|^n$ possible patterns, the environment
contains only a small subset $\{\bX(1),\bX(2),\ldots,\bX(M_c)\}$,
$M_c \ll |\X|^n$. However, before entering the environment, the
system does not know which specific patterns will be present, but
rather knows only their number $M_c$ and that they are generated
according to some probability distribution $p(\x)$.

After being introduced into the environment, the system initially
enters the training phase. During training the system attempts to
form and store an internal representation (memory) of each pattern
along with a semantic label, $w\in\Mc=\{1,2,\ldots,M_c\}$. In
concrete terms, the labels might correspond to a set of actions
the system should undertake when it encounters each pattern,
`pointers' to additional stored information, or `names' for the
patterns. For simplicity, we take the labels to be integers, and
denote the training set by
$\C_x=\{(\bX(1),1),(\bX(2),2),\ldots,(\bX(M_c),M_c)\}$.

\emph{Observations and the testing phase.} After the training
phase, the system enters an `online' \emph{testing} phase. During
testing the observed data is generated as follows. Nature randomly
selects a pattern $W$ according to some distribution $p(w),
w\in\Mc,$ retrieves the corresponding pattern $\x(W)\in\C_x$, and
subjects it to a random transformation $p(\y|\x)$ to produce a
signal $\y=(y_1,y_2,\ldots,y_n)$ with elements in some alphabet
$\Y$. The patterns $\x$ in $\C_x$ thus represent `pure signals' or
prototypes, and the observations $\y\in\Y$ represent distorted and
noise-corrupted variations or \emph{signatures} of the underlying
patterns. The random map $p(\y|\x)$ models two major intrinsic
sources of difficulty in real-world pattern recognition problems:
\emph{signature variation}, differences between the sensory
signals generated on different occasions by the same underlying
object; and \emph{signature ambiguities}, the fact that distinct
objects often produce similar or identical
signatures\footnote{Grenander \cite{grenanderElPatTheory} and
Mumford \cite{Mumford96PABI} have argued that four `universal
transformations' (\emph{noise and blur, superposition, domain
warping}, and \emph{interruptions}) account for most of the
ambiguity and variability in naturally occurring signals.}.

\subsection{Recognition system}
A recognition system consists of three components (functions): A
memory encoder $f$; a sensory encoder $\phi$; and a classifier.
Since we assume the system must be designed prior to insertion
into its environment, the functions $(f,\phi,g)$ must be defined
independent of the specific realizations of the training data
$\C_x$ and sensory data encountered during online operation. On
the other hand, the system design can take account of
\emph{statistical} information about the environment, i.e.
knowledge of the distributions $p(\x)$ and $p(\y|\x)$.

\emph{Encoders.} The memory and sensory encoders $f$ and $\phi$
are mappings from the domains of the raw training and sensory
data, respectively, into some form of approximate \emph{internal
representations}. Encoding may comprise several distinct
operations, such as smoothing and noise reduction, segmentation,
normalization, dimensionality reduction, etc., often collectively
referred to as `feature extraction' procedures
\cite{jain00statistical}. In principle, the role of the resulting
internal data representations may be played by any distinct set of
physical configurations or `states' of the system, provided that
mechanisms exist for associating the training data with these
memory states; inducing appropriate internal states from the
sensory data; and retrieving memorized data, comparing it with
compressed sensory data, and reporting a recognition decision.

Conceptually, we can alternatively regard the internal states of
the system as `codewords,' denoted
$\C_u=\{\u(1),\u(2),\ldots,\u(M_x)\}$ for the memory encoder; and
$\C_v=\{\v(1),\v(2),\ldots,\v(M_y)\}$ for the sensory encoder,
where the codeword alphabets $\U$ and $\V$ are dictated by the
physical nature of the system's memory and sensory systems.

The sensory encoder is then defined as a mapping from the entire
observation space onto the indices $\My=\{1,2,\ldots,M_y\}$ of the
sensory codebook $\phi: \Y^n \ra \My$, $\phi(\y)=\mu$, or
equivalently, onto the codewords $\Cv$. The memory encoder $f$ is
similar, except that it receives labeled inputs and produces
labeled outputs: Given a labeled training pattern $(\x(W),W)$, $f$
associates to it both a memory index $m\in\Mx=\{1,2,\ldots,M_x\}$
and reproduces the class label $w\in\Mc$, representing its storage
in memory. Thus, $f$ is a mapping from the product of the entire
training data space and the set of training labels onto the
product of the memory indices and class labels $f: \X^n \times \Mc
\ra \Mx \times \Mc,$ $f(\x,w)=(m,w)$.

\emph{Classifier.} The classifier, $g$, attempts to infer the
class label of an encountered pattern on the basis of the
compressed sensory information and data stored in memory.
Abstractly, the inference process may take be thought of as a
search through the codebook $\C_u$ for the memory codeword best
matching the current sensory codeword $\v\in\C_v$. Physical
implementations of the matching process may take the form of
computational algorithms; the dynamics of some physical medium
(e.g. a biological neural network); or an abstract decision rule.
Mathematically, a classifier is a mapping $g$ from the encoded
sensory data $\mu=\phi(\y)\in \My$ and the memory data $\C_u$ to a
class label $\hat{w}\in\Mc$, i.e. $g: \My\times \C_u \ra \Mc$,
$g(\mu,\C_u)=\hat{w}$.

\subsection{Figures of merit}
For given distributions $p(\x)$ and $p(\y|\x)$ and data dimension
$n$, there is clearly an intrinsic tradeoff between the number of
internal memory and sensory states, $M_x$ and $M_y$, and the
number of patterns $M_c$ that can be reliably recognized. For our
purposes it is preferable to characterize this tradeoff in a
dimensionless manner, that is, in terms of \emph{rates}. The rates
of the memory and sensory encoders $f$ and $\phi$ are given
respectively by $R_x=\log_2 M_x/n$, $R_y=\log_2 M_y /n$, where
standard interpretations apply (see, e.g.
\cite{cover-thomas-book,csiszar-korner-book,gallager-book}):
Viewing the indices of the memory codebook
$\Mx=\{1,2,\ldots,M_x\}$ as binary strings of length $N_x=\log_2
M_x$, the rate $R_x$ is simply the cost, in bits/symbol, of
representing each $n-$length training pattern $\x\in\X^n$ by a
length-$N_x$ binary string, $R_x=N_x/n$. The analogous
interpretation applies to the sensory codebook. We also quantify
the amount of data in the training set by defining a rate
$R_c=\log_2 M_c /n$, interpreted as the number of training
patterns discriminated \emph{per-symbol} of encoded memory and
sensory data.

\subsection{The meaning of large $n$}
Some of the results below (specifically, the `achievability'
proofs) rely on asymptotic arguments, requiring the parameter $n$
to grow large. Physically, `large-$n$' may correspond to
representing the sensory and memory data at high resolution;
collecting more of it; or making repeated measurements
\cite{jaoNatalia2004}. On the other hand, though our proofs employ
asymptotic arguments, the theorems themselves are stated in terms
of single letter formulas, and in this sense they are independent
of $n$. Hence, the `large-$n$' assumption in the achievability
proofs is not necessarily a fundamental limitation of the theory.

\section{Related issues}\label{sec: related work}
Before formalizing our problem, we briefly comment on some
relationships between the present work and other issues in pattern
recognition.

\emph{Probabilistic modeling.} Our analysis supposes the existence
of probabilistic models for the recognition environment, and that
these distributions are available for use in designing the
recognition system. For some types of random patterns, such as the
pattern of grains on a wooden surface or of magnetic particles on
magnetic tape, estimating the probability distributions is
relatively straightforward \cite{jaoNatalia2004}. Substantial
progress has also been made in modeling more challenging objects,
such as textures in natural imagery
\cite{HeegerBergen_Texture,zhuWuMumford98,Bonet+Viola:1998b,simoncelli00,SrivastavaSimoncelli2003},
and speech signals \cite{jelenik-speechRec-book}. Nevertheless, in
many cases of interest the development of accurate probabilistic
models remains a challenge, and is an active research focus in
pattern recognition research.

\emph{Data compression.} The importance of data compression in
pattern recognition systems appears most clearly articulated in
the neuroscience literature, due largely to the pioneering work of
Horace Barlow. Barlow has written extensively about experimental
evidence and theoretical reasons for believing that principles of
efficient data compression underly the capacity of animal brains
for learning and intelligent behavior (see, e.g.
\cite{barlow61,barlow83,barlow94LargeScale,barlow2001,GardnerMedwin_Barlow2001}).
Additionally, in the past few decades much additional work in
neurobiology has provided experimental evidence for efficient
coding mechanisms in the sensory systems of diverse animals,
including monkeys, cats, frogs, crickets, and flies
\cite{SPIKESbook}. More recently, data compression has come to be
viewed as essential for managing metabolic energy costs in animal
brains \cite{lennie03}.

In the engineering pattern recognition literature, data
compression usually arises in the context of \emph{feature
extraction}. Feature extractors are typically designed with the
objectives of transforming the raw data available to the system
into a format which facilitates easy matching or storage, and is
robust (``invariant'') with respect to characteristic signature
variations in sensory data \cite{lecun-98,jain00statistical}. With
respect to these goals, the volume of data used for internal data
representations is present as an implicit constraint, since
efficient data manipulation is often best achieved by compact
representations. For complex environments, the cost of data
representation becomes critical as an\emph{explicit} design
constraint. Whatever the motivations are, the crucial common
aspect of all data encoding operations for our present purposes is
that they reduce the amount of data available to the system as
compared with the original data (usually in a lossy manner).

\emph{Performance prediction vs. normalization.} Performance
prediction is the problem of characterizing the performance for
specific classes of recognition systems, often with the goal of
discovering the optimal member (e.g. best parameter settings) of a
given class \cite{ProbThPatRec}. By contrast, our objective is to
characterize the requirements for the \emph{existence} of reliable
pattern recognition systems, and to describe absolute performance
limits governing all such systems. In this sense, we aim to
provide normalized performance bounds, with respect to which the
performance of any actual or proposed recognition system may be
evaluated.

\section{Notation} We adopt the following notational conventions. Random
variables are denoted by capital letters (e.g. $U$), and their
values by lowercase letters (e.g. $u$). The alphabet in which a
random variable takes values is denoted by a script capital letter
(e.g. $\mathcal{U}$). Sequences of symbols are denoted either by
boldface letters or with a superscript, interchangeably (e.g.
$\u=u^n=(u_1,u_2,\ldots,u_n)$ denotes a vector which takes values
in the product alphabet $\mathcal{U}^n$). The probability mass
function (p.m.f) for a random variable $U\in \mathcal{U}$ is
denoted by $p_U(u),\;u\in\mathcal{U}.$ When the appropriate
subscript is clear from context, we omit it to simplify notation;
e.g. we usually write $p_U(u)$ simply as $p(u)$. Given random
variables $U,V,W$, we denote the entropy of $U$ by $H(U)$, the
mutual information between $U$ and $V$ by $I(U;V)$, and the
conditional mutual information between $U$ and $V$ given $W$ by
$I(U;V|W)$. The standard acronym `i.i.d' will stand for the phrase
`independent and identically distributed.' To express statements
like `$U$ and $V$ are strongly jointly delta typical' write
$(U,V)\in \T_{UV}$. The definition of strong (delta) joint
typicality will be reviewed in the section where it first appears.
Finally, to express statements like: $X$ and $Z$ are conditionally
independent given $Y$, i.e. $p(x,y,z)=p(y)p(x|y)p(z|y)$, we write
`$X-Y-Z$ form a Markov chain,' or simply $X-Y-Z$.

\section{Formal problem statement}\label{sec:Formal Problem statement}

\begin{definition}
The \emph{environment} for a pattern recognition system, denoted
by
\begin{equation*}
\E=(\Mc,p(w),\X,p(x),p(y|x),\Y),
\end{equation*}
consists of three finite alphabets $\Mc,\X,\Y$, probability
distributions $p(w)$ and $p(x)$ over $\Mc$ and $\X$, and a
collection of probability distributions $p(y|x)$ on $\Y$, one for
each $x\in\X$.
\end{definition}

The interpretations are those given in the preceding section:
$\Mc=\{1,2,\ldots,M_c\}$ is the set of class labels; patterns
vectors are written in the symbols of $\X$; and sensory data
vectors in the symbols of $\Y$. For our analysis we assume:
\begin{itemize}
    \item the distribution over class labels is uniform, $p(w)=1/|\Mc|$ for all $w\in\Mc$;
    \item the pattern components are i.i.d., $p(\x)=\prod_{i=1}^n p(x_i)$;
    \item the observation channel is memoryless, $p(\y|\x)=\prod_{i=1}^n
    p(y_i|x_i)$.
\end{itemize}

\begin{definition}\label{defn:PatRecCode}
An $(M_c,M_x,M_y,n)$ pattern recognition code for an environment
$\E$ consists of three sets of integers
\begin{eqnarray*}
        \Mc&=&\{1,2,\ldots,M_c\} \\
        \Mx&=&\{1,2,\ldots,M_x\} \\
        \My&=&\{1,2,\ldots,M_y\} \\
\end{eqnarray*}
a set of length$-n$ sequences $\bX(i) \in\X^n$,
$i=1,2,\ldots,M_c$, where all components are drawn independently
from $p(x)$ and each sequence is paired with a distinct index from
$\Mc$
\begin{equation*}
    \C_x=\{(\bX(1),1),(\bX(2),2),\ldots,(\bX(M_c),M_c)\};
\end{equation*}
a memory encoder
\begin{eqnarray*}
        f&:& \X^n\times \Mc \ra \Mx\times \Mc; \; f(\x,w)=(m,w);  \\
\end{eqnarray*}
a sensory data encoder
\begin{eqnarray*}
\phi&:& \Y^n \ra \My; \; \phi(\y)=\mu;
\end{eqnarray*}
and a classifier
\begin{equation*}
    g: \My \times \C_u \ra \Mc, \; g(\mu,\C_u)=\hat{w}
\end{equation*}
composed of two submappings $g=g_2\circ g_1$
\begin{eqnarray*}
    g_1 &:& \My \ra \Mx; \; g_1(\mu)=\hat{m} \\
    g_2 &:& \Mx \times \C_u \ra \Mc; \; g_2(\hat{m},\C_u)=\hat{w},
\end{eqnarray*}
where $\C_u$ denotes the encoded training data
\begin{equation*}
\C_u = f(\C_x) = \{(m(1),1),\ldots,(m(M_c),M_c) \}.
\end{equation*}
\end{definition}

For convenience hereafter, we refer to an $(M_c,M_x,M_y,n)$
pattern recognition code by its three constituent mappings
$(f,\phi,g)_n$, or simply as $(f,\phi,g)$ when the integer $n$ is
clear from context.

The rate $\bR=(R_c,R_x,R_y)$ of an $(M_c,M_x,M_y,n)$ code is
\begin{eqnarray*}
R_c &=& \frac{1}{n} \log_2 M_c \\
R_c &=& \frac{1}{n} \log_2 M_x \\
R_c &=& \frac{1}{n} \log_2 M_y, \\
\end{eqnarray*}
where the units are bits per symbol.

For each pattern-label pair $(\x(w),w)\in\C_x$, let $\hat{m}(w)$
be the memory index assigned to $\x(w)$ by the memory encoder $f$,
and let the corresponding sensory data be $\y$. Define two error
events
\begin{eqnarray*}
\ve_1(w) &=& \{\hat{m}\neq m(w)\}   \\
\ve_2(w) &=& \{\hat{w}\neq w\},
\end{eqnarray*}
where $\hat{m}=g_1(\mu)=g_1(\phi(\y))$ and
$\hat{w}=g(\mu,\Cu)=g_2(\hat{m},\Cu)=g_2(g_1(\phi(\y)),\Cu)$; and
denote the union by
\begin{equation*}
\ve(w)=\ve_1(w)\cup \ve_2(w).
\end{equation*}
During the testing phase of operation, if the pattern index
$w\in\Mc$ is selected, let
\begin{equation*}
P_e^n(w)=\text{Pr}\{\ve(w)\}
\end{equation*}
denote the probability of error. Note that these probabilities
depend only on the random vectors $\bX(w)$ and $\bY$ and hence are
determined by the joint distribution $p(\x,\y)=p(\x)p(\y|\x)$. We
define the \emph{average probability of error} of the code as
\begin{equation*}
P_e^n = \frac{1}{M_c} \sum_{w\in\Mc} P_e^n(w).
\end{equation*}
Note that this probability is calculated under a uniform
distribution on the pattern indices, $p(w)=1/M_c$. That is, we
assume that every pattern index $w\in\Mc$, and hence every pattern
$\bX(w)$, is selected with equal probability during the testing
phase.

\begin{comment}\label{Rmk:error}
    Expanding the probability of error in two ways
    \begin{eqnarray*}
        P_e^n &=& Pr\{\ve_1\cup \ve_2 \} \\
        &=& Pr\{ \ve_1 \} + Pr\{ \ve_1^c \}Pr\{ \ve_2|\ve_1^c \} \\
        &=& Pr\{ \ve_2 \} + Pr\{ \ve_2^c \}Pr\{ \ve_1|\ve_2^c \}.
    \end{eqnarray*}
    we see that $P_e^n=0$ if and only if
    \begin{equation*}
    Pr\{\ve_1\} = Pr\{\ve_2\}= Pr\{\ve_1|\ve_2^c\}=
    Pr\{\ve_2|\ve_1^c\}=0.
    \end{equation*}
    The interpretation is that in a reliable pattern recognition system both
    components $g_1$ and $g_2$ of the classifier $g$ must
    function reliably.
\end{comment}

\begin{definition}
    A rate $\bR=(R_x,R_y,R_c)$ is \emph{achievable} in a recognition
    environment $\E$ if for any $\eps>0$
    and for all $n$ sufficiently large, there exists an $(M_c,M_x,M_y,n)$
    code $(f,\phi,g)_n$ with
    \begin{eqnarray*}
        M_c & \geq & 2^{nR_c} \\
        M_x & \leq & 2^{nR_x} \\
        M_y & \leq & 2^{nR_y}
    \end{eqnarray*}
    such that $P_e^n < \eps$.
\end{definition}

\begin{definition}
    The \emph{achievable rate region} $\R$ for a recognition environment $\E$ is the set of
    all achievable rate triples.
\end{definition}

The primary goal of this paper is to characterize the achievable
rate region $\R$ in a way that does not involve the unbounded
parameter $n$, that is, to exhibit a single letter
characterization of $\R$.

\def\yt{{ \tilde{y} }}
\def\Ut{{ \tilde{U} }}
\def\Vt{{ \tilde{V} }}

\section{Main results}\label{Sec: main results}
In this section we present inner and outer bounds on the
achievable rate region $\R$. The bounds are expressed in terms of
sets of `auxiliary' random variable pairs $UV$, defined below. In
these definitions we assume that $U$ and $V$ take values in finite
alphabets $\U$ and $\V$ and have a well defined joint distribution
with the `given' random variables $XY$. To each such pair of
auxiliary random variables $UV$ we associate a set of rates
$\R_{UV}$ defined by
\begin{eqnarray*} \label{eqn:Ruv}
\R_{UV} = \{\bR &:& R_x \geq I(U;X) \\
 &&  R_y \geq I(V;Y) \\
 &&  R_c \leq I(U;V)-I(U;V|X,Y).\}
\end{eqnarray*}
Next, we define two sets of random variable pairs,
\begin{eqnarray*}
\Pi = \{ UV &:& U-X-Y, \\
&& X-Y-V, \\
&& U-(X,Y)-V \}.
\end{eqnarray*}
and
\begin{eqnarray*}
\Po = \{ UV &:& U-X-Y, \\
 && X-Y-V\}. \\
\end{eqnarray*}
When convenient hereafter, we express the three independence
constraints in $\Pi$ as a single `long' Markov chain, $U-X-Y-V$.

Finally, we define two additional sets of rates
\begin{eqnarray*}
\Ri &=&\{\bR: \bR\in \R_{UV} \text{ for some } UV\in \Pi\} \\
\Ro &=&\{\bR: \bR\in \R_{UV} \text{ for some } UV\in \Po\}.
\end{eqnarray*}

\begin{comment}\label{comment: second term in Rc}
Note that for rates in $\Ri$, the long Markov constraint $U-X-Y-V$
implies that the second term in the third inequality of $\R_{UV}$
vanishes, i.e. $I(U;V|XY)=0$.
\end{comment}

Our main results are the following.
\begin{theorem}[Positive theorem: Inner bound]\label{Thm: inner bound}
\begin{equation*}
\Ri\subseteq \R
\end{equation*}
That is, every rate $\bR\in\Ri$ is achievable.
\end{theorem}
\begin{theorem}[Negative theorem: Outer bound]\label{Thm: outer bound}
\begin{equation*}
\Ro\supseteq \R
\end{equation*}
That is, no rate $\bR\notin\Ro$ is achievable.
\end{theorem}
The proofs appear in Appendices \ref{sec:Inner bound proof} and
\ref{Sec: Outer bound proof}.

\begin{remark}\label{noGap}
If either $X=U$ or $Y=V$, or both, then the inner and outer bounds
are identical, since in this case the extra Markov condition
$U-(X,Y)-V$ in the definition of $\Pi$ is automatically satisfied.
For example, if $U=X$, then the condition is equivalent to
$I(U;V|XY)=I(X;V|XY)=0$, which is obviously true. Similar comments
apply if $U$ and $V$ are any deterministic functions of $X$ and
$Y$.
\end{remark}

\section{Discussion of the main results}\label{Sec: discussion}

\subsection{The gap between bounds}
The true achievable rate region is sandwiched between the sets
$\Ri$ and $\Ro$, i.e. $\Ri\subseteq \R \subseteq \Ro$. The gap
between $\Ri$ and $\Ro$ is due to the different independence
constraints in the definitions of $\Pi$ and $\Po$: Whereas
distributions in $\Pi$ satisfy three Markov-chain constraints
$U-X-Y$, $X-Y-V$, and $U-(X,Y)-V$ or, equivalently, the single
`long chain' constraint $U-X-Y-V$, distributions in $\Po$ need
only satisfy the first two `short chain' constraints. Hence, $\Ro$
is the larger rate region and, in general, we expect a gap between
the two regions.

\subsection{Convexity}
One manifestation of the difference between $\Ro$ and $\Ri$ is
that $\Ro$ is convex, while $\Ri$ generally is not. We state this
here as a lemma:

\begin{lemma}\label{lemma: convexity of Ro, nonconvexity of Ri}
$\Ro$ is convex set, in the sense that all rates along the line
connecting any two rates $\bR_1$ and $\bR_2$ contained in $\Ro$
are also contained in $\Ro$.
\end{lemma}
The convexity of $\Ro$ is proved in Appendix \ref{Appdx: Convexity
of Rout}. The nonconvexity of $\Ri$ is apparent from the examples
studied in sections \ref{sec: Binary case} and \ref{sec: Gaussian
case}.

\subsection{Berger's observation and implications}
At least in part, the reason for the gap can be appreciated more
concretely using the following observation made by Berger when
discussing the distributed source coding problem, for which the
currently known inner and outer bounds on the achievable rates are
separated by a similar gap \cite{berger-1977}. Observe that the
long-chain Markov constraint on $\Pi$ implies that each
corresponding joint distribution over $UV$ given $XY$ must
factorize into a product of marginal distributions,
$p(u,v|x,y)=p(u|x)p(v|y)$. By contrast, the less restrictive
constraints on $\Po$ admit pairs whose joint distributions are
convex mixtures of product marginals; that is, distributions of
the form
\begin{equation*}
p(uv|xy)=\sum_{q\in\Q} p(q)p(u|x,q)p(y|v,q).
\end{equation*}
More explicitly, we can represent the set of all such auxiliary
random variable pairs as follows.

\begin{definition}
Let
\begin{equation*}
\Pm=\{UV: U=(U_Q,Q),\; V=(V_Q,Q) \},
\end{equation*}
where $Q$ is any discrete random variable with a finite alphabet
$\Q$ which is independent of $X$ and $Y$, and for each $q\in\Q$
the pair $U_q V_q \in\Pi$.
\end{definition}

Clearly, there is potentially a much larger set of distributions
for $UV$ pairs in $\Pm$ than in $\Pi$.

However, while $\Pm$ is clearly contained in $\Po$, it is unknown
whether or under what conditions $\Pm=\Po$. Further, if we define
the additional rate region
\begin{equation*}
\Rm = \{\bR: \bR\in\R_{UV} \text{ for some } UV\in\Pm \},
\end{equation*}
and let $\Co(\Ri)$ denote the convex hull of $\Ri$
\begin{equation*}
\Co(\Ri) = \{\bR: \bR=\theta \bR_1 +\tb\bR_2, \; \bR_1,\bR_2
\in\Ri,\; 0\leq \t \leq 1\}
\end{equation*}
where $\tb=1-\t$, then it is easy to verify that the following
logical statement holds:
\begin{eqnarray}\label{eqn: equivalences}
\text{If } && \Po = \Pm \\ \nonumber \text{then} && \Ro = \Rm =
\Co(\Ri).
\end{eqnarray}
Thus, it is unknown whether the presence of mixture distributions
in $\Po$ is enough to account for all of the gap between $\Ri$ and
$\Ro$. As discussed below in subsection \ref{Sec: closing the
gap}, (\ref{eqn: equivalences}) has interesting implications for
closing the gap.

\subsection{Relationship with distributed source coding}
\begin{figure}[htbp]
\begin{center}
    \psfrag{pxy}{$p(\x,\y)$}
    \psfrag{phi}{$\phi(\cdot)$}
    \psfrag{f}{$f(\cdot)$}
    \psfrag{x}{$\bX$} \psfrag{y}{$\bY$}
    \psfrag{m}{$m$} \psfrag{mu}{$\mu$}
    \psfrag{g}{$g(\cdot,\cdot)$}
    \psfrag{xy}{$(\hat{\bX},\hat{\bY})$}
    \includegraphics[width=2.5in]{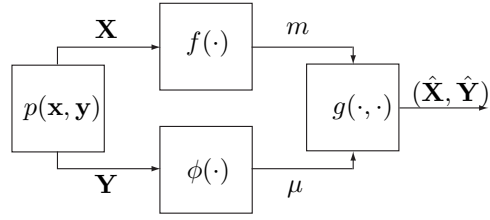}
    \caption{The distributed source coding problem.}
    \label{Fig:DSC}
\end{center}
\end{figure}

Some interesting connections hold between the results of Tung and
Berger \cite{Tung78Thesis,berger77MultTermSC} for the distributed
source coding (DSC) problem and our results in theorems \ref{Thm:
inner bound} and \ref{Thm: outer bound}. Briefly, the situation
treated in the DSC problem, diagrammed in figure \ref{Fig:DSC}, is
as follows. Two correlated sequences, $\bX$ and $\bY$, are encoded
separately as $m=f(\bX)$, $\mu=\phi(\bY)$, and the decoder $g$
must reproduce the original sequences subject to a fidelity
constraint, $(E d_x(\hat{\bX},\bX), E d_y(\hat{\bY},\bY)) \leq
\bD$, whrere $\bD=(D_x,D_y)$. The problem is to characterize, for
any given distortion $\bD$, the set of achieveable rates
$\R(\bD)$.

%\input{Fig_DSCblockDiagram}
%\graphicspath{{MatlabFigures/IllustratorPics/}}

The known inner and outer bounds for the DSC problem are as
follows. Let $\Pi$ and $\Po$, be defined as above, and define two
new sets incorporating the distortion constraint
\begin{eqnarray*}
    \Pi(\bD)&=&\Pi\cap \P_{UV}(\bD) \\
    \Po(\bD)&=&\Po\cap \P_{UV}(\bD),
\end{eqnarray*}
where
\begin{equation*}
\P_{UV}(\bD)=\{UV: \; \exists \; \hat{X}(U,V),
     \hat{Y}(U,V) \text{ s.t. } (E d_x(\hat{X},X), E d_y(\hat{Y},Y)) \leq
    \bD \}.
\end{equation*}
Parallelling equation \ref{eqn:Ruv}, also define the sets of rates
\begin{eqnarray*}
\bar{\R}_{UV} = \{\bR &:& R_x \geq I(U;X|V) \\
 &&  R_y \geq I(V;Y|U) \\
 &&  R_x+R_y \geq I(UV;XY).\}
\end{eqnarray*}
and
\begin{eqnarray*}
\Ri(\bD) &=&\{\bR: \bR\in \bar{\R}_{UV} \text{ for some } UV\in \Pi(\bD)\} \\
\Ro(\bD) &=&\{\bR: \bR\in \bar{\R}_{UV} \text{ for some } UV\in
\Po(\bD)\}.
\end{eqnarray*}
Then the Berger-Tung bounds for the DSC problem can be expressed
as $\Ri(\bD)\subseteq \R(\bD)$, and $\Ro(\bD)\supseteq \R(\bD)$.

With the results presented in this way, the formal similarities
between our pattern recognition problem and the DSC problem are
obvious. Additionally, ignoring the distortion constraints for the
moment, the pattern recognition problem can be thought of as a
kind of generalization of the DSC problem, with the added
complication that the `decoder' receives not one sequence $\bX$
but $M_c=2^{nR_c}$ such sequences, and must first determine which
is the appropriate one with which to jointly decode the second
received sequence $\bY$. This extra discrimination evidently
requires extra information to be included at the encoders. This
`rate excess' is the difference between the minimum encoding rates
required for the DSC and pattern recognition problems. Using the
the short-chain Markov constraints $U-X-Y$ and $X-Y-V$, the rate
excess for the $\bX$ encoder is
\begin{eqnarray*}
I(X;U)-I(X;U|V) &=& I(X;U)-I(XY;U|V) \\
&=& I(X;U)-[I(XY;UV)-I(XY;V)] \\
&=& I(X;U)+I(Y;V)-I(XY;UV) \\
&=& I(U;V)-I(U;V|XY)
\end{eqnarray*}
and, by symmetry, at the $\bY$ encoder the excess required rate is
\begin{equation*}
I(Y;V)-I(Y;V|U) = I(U;V)-I(U;V|XY).
\end{equation*}
Thus, the excess rate required at either terminal is directly
related to the maximum number of patterns that must be
discriminated, $M_c=2^{nR_c}$, $R_c = I(U;V)-I(U;V|XY)$.

\subsection{Extension of the inner bound}
The following results provide a way to reduce the gap between
$\Ri$ and $\Ro$ `from below,' by improving on the inner bound.

\begin{theorem}\label{theorem: extension}
If the point $\bR=(R_c, R_x, R_y)$ is achievable, then for any $0
< \theta \le 1$, the point $\bR'=\theta \bR$ is achievable.
\end{theorem}
\begin{corollary}\label{corollary: extension corollary}
Let
\begin{equation*}
\R' = \{\bR: \bR=\theta \bR', \; \bR'\in\Ri,\; 0\leq \theta\leq
1\}.
\end{equation*}
Then $\R' \subseteq \R$.
\end{corollary}

The theorem and corollary are proved in Appendix \ref{Appendix:
proof of extension}. As discussed in the next subsection, this
extension of the inner bound may in some cases allow us to close
the gap, i.e. in cases where the expression for the convex hull of
$\Ri$ simplifies such that $\Co(\Ri)=\R'$. Specific examples where
this appears to be the case include the binary and Gaussian
examples discussed in sections \ref{sec: Binary case} and
\ref{sec: Gaussian case}.

\subsection{On closing the gap}\label{Sec: closing the gap}
What additional results would be needed to determine the true
achievable rate region $\R$? To explore this question, consider
the following hypothetical statements and their implications.
\begin{description}
    \item[(a)] $\Po=\Pm$
    \item[(b)] $\Co(\Ri) = \R'$
    \item[(c)] $\R$ is convex
    \item[(d)] $\R = \Ro$
\end{description}
We emphasize that none of these statements have been proven.
Nevertheless, the following Lemmas, stated in `if-then' form, are
true.
\begin{lemma}\label{desc: star}
(a),(b) $\Rightarrow$ (d)
\end{lemma}
\begin{lemma}\label{desc: starstar}
(a),(c) $\Rightarrow$ (d)
\end{lemma}

The proof of Lemma \ref{desc: star} is as follows. Assuming
$\Co(\Ri)=\R'$, then by corollary \ref{corollary: extension
corollary} the convex hull is achievable, $\Co(\Ri)\subseteq \R$.
But by (\ref{eqn: equivalences}) our assumption (a) implies $\Ro =
\Rm = \Co(\Ri)$, hence $\Ro\subseteq \R$. Combining this with
theorem \ref{Thm: outer bound} we have $\Ro\subseteq \R$ and
$\Ro\supseteq \R$, or $\R=\Ro$.

Lemma \ref{desc: starstar} follows from straightforward
timesharing arguments, as shown in Appendix \ref{appendix:
starstar proof}.

Both Lemmas \ref{desc: star} and \ref{desc: starstar} suggest
potential routes for establishing the true achievable rate region
$\R$ by expanding the inner bound $\Ri$. While we expect that
premises (a) and (b) hold in certain cases, we suspect that they
are not true in general; we have no current guess about (c). On
the other hand, if $\Ro$ is larger than $\Rm$, then it may still
be possible to establish the true achievable rate region $\R$ by
tightening the outer bound, possibly down to $\Rm$. Thus $\Ro$ and
$\Rm$ are presently the most promising candidates for $\R$.

\subsection{Degenerate cases}
We now briefly examine the degenerate cases where either $X=U$, or
$Y=V$, or both. These simple cases have clear interpretations and
are thus useful for building intuition about the general results
of theorems \ref{Thm: inner bound} and \ref{Thm: outer bound}.
Note that in these cases $I(U;V|XY)=0$, hence the third inequality
in the definition of $\R_{UV}$ \ref{eqn:Ruv} simplifies to $R_c
\leq I(U;V)$. Additionally, in these cases there is no gap, i.e.
the inner and outer bounds are \emph{equal}; see Remark
\ref{noGap}.

\emph{Unlimited senses and memory.} First, consider a system in
which the budgets for memory and sensory representations are
unrestricted, i.e. no compression is required. In this case, we
can effectively treat the memories and sensory representations as
if they were veridical; i.e. we can set $U=X$ and $V=Y$. The
theorem constraints then become $R_x\geq I(X;X)=H(X),\;R_y \geq
I(Y;Y)=H(Y)$, and
\begin{equation}\label{no limits}
    R_c \leq I(U;V)=I(X;Y).
\end{equation}
This result indicates that, in the absence of compression, the
recognition problem is formally equivalent to the following
classical communication problem: Transmit one of $M_c=2^{nR_c}$
possible messages (patterns) to a receiver (the recognition
module) \cite{jaoNatalia2004}. In this case, the objects can be
thought of as codewords which are stored without compression for
direct comparison with the sensory data. This is the setup of the
random coding proof of Shannon's channel coding theorem, which
gives the rates at which reliable communication is possible as
those below the mutual information between the source (analogous
to the memory here) and the received signals, $I(X;Y)$
\cite{cover-thomas-book,shannon-1948}. This is exactly the
condition expressed by (\ref{no limits}). The condition specifies
an upper bound on the number of objects the system may be trained
to recognize through the relation $M_c=2^{nR_c}$.

\emph{Unlimited memory, limited senses.} Next, suppose that memory
is effectively unlimited, so that we can put $U=X$, but sensory
data may be compressed. In this case, we can readily rewrite the
condition on $R_c$ as
\begin{equation}\label{ltd senses}
R_c \leq I(X;Y)-I(X;Y|V).
\end{equation}
We check the extreme cases: If $Y$ is fully informative about $V$,
$Y=\phi^{-1}(V)$, then $I(X;Y|V)=H(Y|V)-H(Y|X,V)=0$, and we
recover the case discussed above. For intermediate cases where $V$
is partially informative, then the effect of $V$ is to degrade the
achievable performance of the system below that possible with
`perfect senses,' and the reduction incurred is $I(X;Y|V)$. In the
extreme case that $V$ is utterly uninformative (i.e. independent
of $Y$), then $I(X;Y|V)=I(X;Y)$, and we get $R_c = 0$, or $M_c\leq
2^{nR_c}=1$, hence the system is useless.

\emph{Limited memory, unlimited senses.} In the case of limited
memory but unrestricted resources for sensory data representation,
we get an expression symmetric with the previous case:
\begin{equation}
R_c \leq I(X;Y)-I(X;Y|U).
\end{equation}
As before, if the memory is perfect ($U=X$), we get
$I(X;Y|U)=I(X;Y|X)=0$, recovering the channel coding constraint
$R_c \leq I(X;Y)$; assuming useless memories yields $R_c \leq
I(X;Y)-I(X;Y)=0$; and intermediate cases place the system between
these extremes.

\subsection{Rate region surfaces}\label{subsec: surface
expressions} An equivalent way to characterize the sets $\R$,$\Ri$
and $\Ro$ that will be useful in sections \ref{sec: Binary case}
and \ref{sec: Gaussian case} is to specify the boundary or
\emph{surface} of each region. For $\R$, the surface is
\begin{eqnarray*}
r(r_x,r_y) &=& \max_{R\in \C(r_x,r_y)} R_c, \text{ where} \\
\C(r_x,r_y)&=& \{\bR : \bR\in\R, \; R_x=r_x,\; R_y=r_y\}.
\end{eqnarray*}

Similarly, by direct extension of theorems \ref{Thm: inner bound}
and \ref{Thm: outer bound}, the surfaces of $\Ri$ and $\Ro$ are
specified by
\begin{eqnarray}\label{eqn:surfaces}
r_{in}(r_x,r_y) &=& \max_{UV \in \C_{in} (r_x,r_y)} I(U;V)-I(U;V|XY) \\
r_{out}(r_x,r_y) &=& \max_{UV \in \C_{out} (r_x,r_y)}
I(U;V)-I(U;V|XY) \nonumber,
\end{eqnarray}
where
\begin{eqnarray*}
    \C_{in}(r_x,r_y)&=&\{UV\in\Pi : \; r_x \geq I(U;X),\; r_y \geq
    I(V;Y)\}\\
    \C_{out}(r_x,r_y)&=&\{UV\in\Po :\; r_x \geq I(U;X),\; r_y \geq
    I(V;Y)\}.
\end{eqnarray*}
A useful alternative form comes from rewriting the right hand side
of (\ref{eqn:surfaces}) as
\begin{equation} \label{eqn: alt form}
\begin{split}
I&{}(U;V)-I(U;V|XY)\\
&{} = I(U;V)-H(U|XY)-H(V|XY)+H(UV|XY)  \\
&{} = I(U;V)-H(U|X)-H(V|Y)+H(UV|XY) \\
&{} = I(X;U)+I(Y;V)-I(XY;UV).
\end{split}
\end{equation}
The second line follows from the Markov constraints $U-X-Y$ and
$X-Y-V$. Hence,
\begin{equation}\label{eqn:surfaces2}
r_{*}(r_x,r_y) = \max \; I(X;U)+I(Y;V) -I(XY;UV) \\
\end{equation}
where the subscript $*$ stands for $\text{in}$ or $\text{out}\}$
and the maximization is over $\C_{in}(r_x,r_y)$ or
$\C_{out}(r_x,r_y)$, respectively.

In what follows we seek \emph{explicit} formulas for
$\ri(r_x,r_y)$ and $\ro(r_x,r_y)$, which do not involve the
optimization over the sets $\C_{in}(r_x,r_y)$ and
$\C_{out}(r_x,r_y)$.

\section{Binary case}\label{sec: Binary case} In this section we study a simple case in which the
alphabets for the training patterns and sensory data are binary,
$\X=\Y=\{0,1\}$. The training patterns $\bX=(X_1,\ldots,X_n)$ are
generated by $n-$independent drawings from a uniform Bernoulli
distribution, $X\sim B(1/2)$. Observations $\bY=(Y_1,\ldots,Y_n)$
are outputs of a binary symmetric channel with crossover
probability $q$
\begin{equation*}
p(y|x)  =\left(%
\begin{array}{cc}
  \qb & q \\
  q & \qb \\
\end{array}%
\right),
\end{equation*}
where $\qb = 1-q$. Equivalently, we can represent $Y$ as
$Y=X\oplus W,$ where $W\sim B(q)$ and is independent of $X$.

We now propose explicit formulas for $r_{in}(r_x,r_y)$ and
$r_{out}(r_x,r_y)$ in this binary case. Our formulas involve the
following two functions. First, define
\begin{equation*}
g(r_x,r_y)=1-h(q*q_x*q_y),
\end{equation*}
where $q_x$ and $q_y$ are specified implicitly by
\begin{eqnarray*}
r_x&=&1-h(q_x) \\
r_y&=&1-h(q_y);
\end{eqnarray*}
$h(\cdot)$ is the binary entropy function
$h(x)=-x\log(x)-(1-x)\log(1-x)$; and $q_x,q_y\in [0,1/2]$ to
ensure that $h(\cdot)$ is invertible.  Next, let $g^*(r_x,r_y)$
denote the \emph{upper concave envelope} of $g(r_x,r_y)$,
\begin{equation*}
g^*(r_x,r_y)=\sup \;\; \theta
g(r_{x_1},r_{y_1})+\bar{\theta}g(r_{x_2},r_{y_2}),
\end{equation*}
where $\tb=1-\theta$. The supremum is over all combinations
$(\theta,r_{x_1},r_{y_1},r_{x_2},r_{y_2})$ such that
\begin{equation*}
(r_x,r_y) = \theta(r_{x_1},r_{y_1})+\tb(r_{x_2},r_{y_2}),
\end{equation*}
and each variable in the optimization is restricted to the unit
interval $[0,1]$. As explained in Appendix \ref{appendix: Convex
hulls}, in both the binary case and the corresponding Gaussian
case considered in the next section, the expression for the convex
hull of the inner bound simplifies to
\begin{equation*}
g^*(r_x,r_y)=\sup \;\; \theta g(r_{x}',r_{y}'),
\end{equation*}
and the supremum is over all combinations $(\theta,r_{x}',r_{y}')$
such that
\begin{equation*}
(r_x,r_y) = \theta(r_{x}',r_{y}').
\end{equation*}

\begin{conjecture}
    In the binary case the surfaces of $\Ri$ and $\Ro$
    are
    \begin{eqnarray*}
        \ri(r_x,r_y) &=& g(r_x,r_y) \\
        \ro(r_x,r_y) &=& g^*(r_x,r_y).
    \end{eqnarray*}
\end{conjecture}

%\input{Fig_BinarySurfaces}
%\graphicspath{{MatlabFigures/BinaryCase/}}
\begin{figure}[htbp]
\begin{center}
    \psfrag{rx}{$r_x$} \psfrag{ry}{$r_y$} \psfrag{z}{$z$}
    \psfrag{0}{$0$} \psfrag{1}{$1$} \psfrag{2}{$2$}
    \psfrag{3}{$3$} \psfrag{5}{$5$} \psfrag{.}{$.$}
    \psfrag{8}{$8$}
    \subfigure[]{{\includegraphics[width=1.5in]{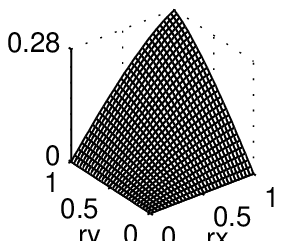}}}
    \subfigure[]{{\includegraphics[width=1.5in]{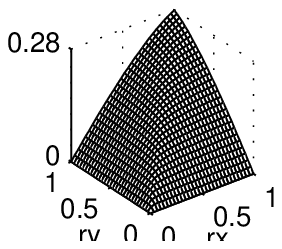}}}
    \\
    \subfigure[]{{\includegraphics[width=1.5in]{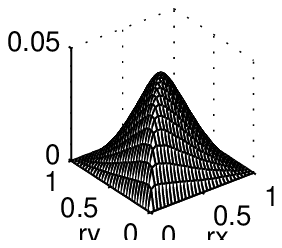}}}

    \caption{Surfaces of the binary inner bound $z=r_{in}$ (a) and outer bound $z=r_{out}$ (b) regions;
    and difference between the outer bound and inner
    bounds $z=r_{in}-r_{out}$ (c).  In these plots the crossover probability $q = 0.2$.}
    \label{Fig:BinarySurfaces}
\end{center}
\end{figure}

From Theorem \ref{theorem: extension}, $g^*(r_x,r_y)$ is in fact
achievable. Thus, if the conjecture on the outer bound is true,
then there is no gap between the inner and outer bounds, and
$g^*(r_x,r_y)$ defines the achievable rate region. Figure
\ref{Fig:BinarySurfaces} shows the inner and outer bounds and
their difference.

To establish these conjectures we must prove both the `forward'
inequalities $\ri \geq g$, $\ro\geq g^*$, and the `backward'
inequalities $\ri \leq g$, $\ro \leq g^*$. The backward
inequalities remain to be proven, whereas the forward inequalities
can be proven by relatively straightforward constructions, as we
now show.

\begin{proof}($\ri(r_x,r_y)\geq g(r_x,r_y)$)
Let $W_x\sim B(q_x)$, $W_y\sim B(q_y)$ be binary random variables
independent of $X$ and $Y$, and define
\begin{eqnarray*}
    U&=& X\oplus W_x \\
    V&=& Y\oplus W_y.
\end{eqnarray*}
The pair $UV$ is obviously in $\Pi$. Furthermore,
    \begin{eqnarray*}
        I(X;U) &=& H(X)-H(X|U) \\
        &=& 1-H(U\oplus W_x|U) \\
        &=& 1- H(W_x) \\
        &=& 1-h(q_x), \\
        I(V;Y) &=& H(Y)-H(Y|V) \\
        &=& 1-H(V\oplus W_y|V) \\
        &=& 1-H(W_y |V) \\
        &=& 1-h(q_y), \\
        I(U;V) &=& H(V)-H(V|U) \\
        &=& 1-H(U\oplus W_x\oplus W\oplus W_y|U) \\
        &=& 1-h(q_x*q*q_y). \\
    \end{eqnarray*}
    Setting $r_x=I(U;X)=1-h(q_x)$, and $r_y=I(Y;V)=1-h(q_y)$, we have $UV\in\Pi$
    and $UV\in\C(r_x,r_y)$. Hence,
    \begin{equation*}
    r_{in}(r_x,r_y)=\max_{UV\in\C(r_x,r_y)}I(U;V)\geq 1-h(q*q_x*q_y)=g(r_x,r_y).
    \end{equation*}
\end{proof}

\begin{proof}($\ro(r_x,r_y)\geq g^*(r_x,r_y)$)
Using the same construction as in the forward proof for the inner
bound formula, define two pairs of random variables $(U_1
V_1),(U_2 V_2) \in \Pi\subseteq \Po$ such that
\begin{eqnarray*}
r_{x_1}&=& I(U_1;X)=1-h(q_{x_1}), \\
r_{y_1}&=& I(V_1;Y)=1-h(q_{y_1}), \\
r_{x_2}&=& I(U_2;X)=1-h(q_{x_2}), \\
r_{y_2}&=& I(V_2;Y)=1-h(q_{y_2}). \\
\end{eqnarray*}
Let $(r_x,r_y)=\theta(r_{x_1},r_{y_1})+\tb (r_{x_2},r_{y_2})$,
$\theta\in[0,1]$. Since $r_{out}(r_x,r_y)$ is convex, we have
\begin{eqnarray*}
\ro(r_x,r_y) &\geq& \theta r_{out}(r_{x_1},r_{y_1}) +\tb \ro(r_{x_2},r_{y_2}) \\
&\geq& \theta g(r_{x_1},r_{y_1})+\tb g(r_{x_2},r_{y_2}).
\end{eqnarray*}
The inequalities above hold for all valid choices of
$\theta,r_{x_1},r_{x_2},r_{y_1},r_{y_2}$, hence
$r_{out}(r_x,r_y)\geq g^*(r_x,r_y)$, as desired.
\end{proof}

\section{Gaussian case}\label{sec: Gaussian case} We now consider a Gaussian version of our problem.
Let $X$ and $Y$ be zero-mean Gaussian random variables with
correlation coefficient $\rho_{xy}$. In parallel with our
discussion of the binary case, we propose explicit formulas for
the surfaces of $\Ri$ and $\Ro$ for the Gaussian case, this time
in terms of the following two functions. In both formulas, let
\begin{eqnarray*}
r_x &=& -\frac{1}{2}\log(1-\rxu^2) \\
r_y &=& -\frac{1}{2}\log(1-\ryv^2).
\end{eqnarray*}
Note that these expressions determine the correlation coefficients
$\rxu$ and $\ryv$. Define
\begin{equation}
G(r_x,r_y)  = -\frac{1}{2}\log(1-\rxy^2\ryv^2\rxu^2).
\end{equation}
and
\begin{equation}
G^*(r_x,r_y) = r_x+r_y +\frac{1}{2}\log[1+\frac{2\rho^2 \gamma
-\beta}{1-\rho^2}],
\end{equation}
where
\begin{eqnarray}\label{eqn: gamma and beta}
\gamma &=& \rxy\rxu\ryv, \\
\beta &=& \rxu^2+\ryv^2-(1-\rxy^2)\rxu^2\ryv^2,  \nonumber \\
\rho &=& \frac{\beta}{2\gamma} -
\sqrt{\left(\frac{\beta}{2\gamma}\right)^2-1}. \nonumber
\end{eqnarray}

\begin{conjecture}
    In the Gaussian case the surfaces of $\Ri$ and $\Ro$
    are
    \begin{eqnarray*}
        \ri(r_x,r_y) &=& G(r_x,r_y) \\
        \ro(r_x,r_y) &=& G^*(r_x,r_y).
    \end{eqnarray*}
\end{conjecture}
Figure \ref{Fig:GaussSurfaces} shows plots of the inner and outer
bounds and their difference, as well as the difference between the
outer bound and the convex hull of the inner bound. Interestingly,
unlike the binary case, for the Gaussian case the outer bound is
not equal to the convex hull of the inner bound.

The following proof relies on some basic properties of the mutual
information between Gaussian random variables, given as Lemmas in
Appendix \ref{appendix: Gaussian properties}.

%\input{Fig_GaussSurfaces}
%\graphicspath{{MatlabFigures/GaussianCase/}}
\begin{figure}[htbp]
\begin{center}
    \psfrag{rx}{$r_x$} \psfrag{ry}{$r_y$} \psfrag{z}{$z$}
    \psfrag{0}{$0$} \psfrag{1}{$1$} \psfrag{2}{$2$}
    \psfrag{3}{$3$} \psfrag{5}{$5$} \psfrag{.}{$.$}
    \subfigure[]{{\includegraphics[width=1.5in]{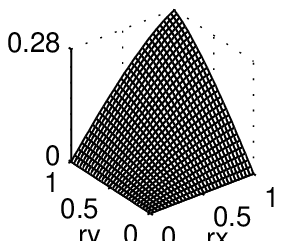}}}
    \subfigure[]{{\includegraphics[width=1.5in]{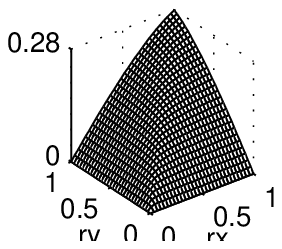}}}
    \\
    \subfigure[]{{\includegraphics[width=1.5in]{Fig_GaussDiff_OBIB.eps}}}
    \subfigure[]{{\includegraphics[width=1.5in]{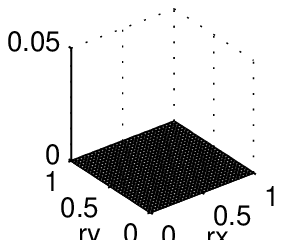}}}

    \caption{Surfaces of the Gaussian inner bound $z=r_{in}$ (a) and outer bound $z=r_{out}$ (b) regions;
    and differences between the outer bound and inner
    bounds $z=r_{in}-r_{out}$ (c) and between the outer bound and the convex
    hull of the inner bound $z=r_{out}-\calH(r_{in})$ (d). In these plots $\rxy = 0.8$.}
    \label{Fig:GaussSurfaces}
\end{center}
\end{figure}

In the analysis that follows, we assume that the true
distributions are Gaussian.  Under this assumption, we solve the
inner and outer bounds.  If the true distributions are Gaussian,
then our conjecture is true.

\begin{proof}($\ri(r_x,r_y) = G(r_x,r_y)$)
As noted in Appendix \ref{appendix: Gaussian properties}, mutual
informations between jointly Gaussian random variables are
completely determined by their correlation coefficients. For a
length-4 Markov chain $U-X-Y-V$ of jointly Gaussian random
variables $I(U;V|XY)=0$ and, applying Lemma \ref{lemma:
gauss-markov} from Appendix \ref{appendix: Gaussian properties} we
have $\ruv=\rxu\rxy\ryv$, hence
\begin{eqnarray*}
I(U;V)-I(U;V|XY) = -\frac{1}{2}\log(1-\rxu^2\rxy^2\ryv^2).
\end{eqnarray*}
This mutual information is maximized when the constraints
$I(X;U)\leq r_x$, $I(Y;V)\leq r_y$ are satisfied with equality,
hence when $\rxu$ and $\ryv$ satisfy $r_x =
-\frac{1}{2}\log(1-\rxu^2)$ and $r_y =
-\frac{1}{2}\log(1-\ryv^2)$. This proves the theorem.
\end{proof}

The following proof for the surface of the outer bound region uses
the form of $\ro(r_x,r_y)$ in (\ref{eqn:surfaces2}). We assume
that the constraints on $r_x$ and $r_y$ are satisfied with
equality. In this case, the optimization problem reduces to that
of minimizing the $I(XY;UV)$ subject to the length-3 Markov
constraints $U-X-Y$, $X-Y-V$.
\begin{proof}($\ro(r_x,r_y) = G^*(r_x,r_y)$)
Using Lemma \ref{lemma: length four correlations} from appendix
\ref{appendix: Gaussian properties}, we have
\begin{eqnarray*}
C_{xy,uv} &=&
\left[%
\begin{array}{cc}
  \rxu & \rxv \\
  \ryu & \ryv \\
\end{array}%
\right] = \left[%
\begin{array}{cc}
  1 & \rxy \\
  \rxy & 1 \\
\end{array}%
\right]
\left[%
\begin{array}{cc}
  \rxu & 0 \\
  0 & \rxu \\
\end{array}%
\right].
\end{eqnarray*}
The left hand matrix in this decomposition is $C_{xy,xy}$, denoted
hereafter simply as $C$, and we denote the righthand matrix by
$D$.  Then applying Lemma \ref{lemma: Gaussian info} from appendix
\ref{appendix: Gaussian properties} yields
\begin{equation*}
\begin{split}
&{}I(XY;UV) \\
&{} =
\frac{1}{2}\log|C|-\frac{1}{2}\log|C-C_{xy,uv}C^{-1}_{uv,uv}C_{uv,yx}|
\\
&{}= \frac{1}{2} \log|C|-\frac{1}{2}\log|C-CD C^{-1}_{uv,uv} D C| \\
&{}= -\frac{1}{2} \log|C|-\frac{1}{2}\log|C^{-1} -D C^{-1}_{uv,uv}
D|.
\end{split}
\end{equation*}
Substituting for the $2\times 2$ matrices in this last expression
and rearranging terms yields
\begin{equation*}
I(XY;UV) = -\frac{1}{2}\log[1+\frac{2\ruv^2 \gamma
-\beta}{1-\ruv^2}],
\end{equation*}
where $\gamma$ and $\beta$ are defined in (\ref{eqn: gamma and
beta}).

By assumption, $\rxu$ and $\ryv$ are being held fixed, so we are
optimizing $I(XY;UV)$ only with respect to $\ruv$. Setting
$\partial I(XY;UV) /\partial \ruv=0$ and solving, we obtain that,
if $\beta >2\gamma>0$, then the maximum is achieved at $\ruv^* =
\rho$, where $\rho$ is defined in (\ref{eqn: gamma and beta}).

To complete the proof we must show that $\beta>2\gamma>0$. Noting
that $\beta,\gamma>0$ and substituting, the desired inequality
becomes
\begin{equation*}
\rxu^2 + \ryv^2 -\rxu^2\ryv^2 > 2\rxy\rxu\ryv-\rxy^2\rxu^2\ryv^2.
\end{equation*}
Subtracting 1 from each side and factoring yields the equivalent
inequality
\begin{equation*}
-(1-\rxu^2)(1-\ryv^2) > -(1-\rxy\rxu\ryv)^2.
\end{equation*}
To show that this holds for all $\rxy$, note that the maximum of
the right hand side is achieved by $\rxy=1$, so that the
inequality becomes
\begin{equation*}
(1-\rxu^2)(1-\ryv^2)-(1-\rxu\ryv)^2<0.
\end{equation*}
This inequality holds, since
\begin{equation*}
\begin{split}
(1-&{}\rxu^2)(1-\ryv^2)-(1-\rxu\ryv)^2 \\
&{} 1-\ryv^2-\rxu^2+\rxu^2\ryv^2 - [1-2\rxu\ryv+\rxu^2\ryv^2] \\
&{}= -\rxu^2-\ryv^2+2\rxu\ryv \\
&{}= (\rxu-\ryv)(\ryv-\rxu) \\
&{}= -(\rxu-\ryv)^2 \\
&{}< 0.
\end{split}
\end{equation*}
\end{proof}

\appendices
\section{Proof of the inner bound}\label{sec:Inner bound proof} In this section
we prove the inner bound $\Ri\subseteq \R$, theorem \ref{Thm:
inner bound}. The proof relies on standard random coding arguments
and properties of strongly jointly typical sets
\cite{cover-thomas-book}. Given a joint distribution $p(xyuv)$,
the strongly jointly $\d$-typical set is defined by
\begin{equation*}
\T^\d_{UVXY} = \left\{ \x\y\u\v:
\left|\frac{N(xyuv|\x\y\u\v)}{n}-p(xyuv)\right| \leq \d \; \forall
xyuv\in\X\Y\U\V\right\},
\end{equation*}
where $N(xyuv|\x\y\u\v)$ is the number of times the symbol
combination $xyuv$ occurs in $\x\y\u\v$. Likewise, we write e.g.
$\T^\d_X$, $\T^\d_{XY}$, $\T^\d_{XYU}$ for singles, pairs, and
triples. We will also use conditionally strongly jointly
$\d$-typical sets, for example
\begin{equation*}
\T^\d_{\x U}= \{\u: (\x\u)\in \T^\d_{XU}\}.
\end{equation*}
The subscripts are omitted when context allows. We will also need
the fact that for any positive numbers $\d,\eps>0$, fixed vector
$\x$, and large enough $n$,
\begin{equation}\label{Eqn: TypicalSetProp}
2^{-n[I(X;Y)+\eps]} \leq Pr(\x\bY\in\T^\d_{\x Y}) \leq
2^{-n[I(X;Y)-\eps]}.
\end{equation}

\begin{proof}
To begin, let $\bR=(R_c,R_x,R_y)$ be any rate triple in $\Ri$, and
let $\eps>0$ be any positive constant. Then there exists a pair of
random variables $UV\in\Pi$ such that $\bR\in\R_{UV}$. We wish to
prove $\bR\in\R$. To this end, we will use $UV$ to construct an
$(M_x,M_y,M_c,n)$ pattern recognition code $(f,\phi,g)$, with $M_c
\geq 2^{nR_c}$, $M_x \leq 2^{nR_x}$, and $M_y \leq 2^{nR_y}$, such
that $P_e^n \leq \eps$ for a sufficiently large integer $n$.

For concreteness, we will suppose that the mappings $f$, $\phi$
and $g$ are implemented in distinct memory, sensory, and
recognition `modules,' respectively, each of which `knows' the
joint distribution $p(xyuv)$.

\textbf{Random codebook generation.} To serve as codewords, select
$M_x$ length$-n$ vectors by sampling with replacement from a
uniform distribution over the set $\T^\d_U$. Assign each codeword
a unique index $i\in\Mx$, where $\Mx=\{1,2,\ldots,M_x\}.$ Denote
the resulting codebook
\begin{equation*}
        \B_u = \{\u(1),\u(2),\ldots, \u(M_x)\},
\end{equation*}
where the $\u(i)$ are the indexed codewords.

Similarly, for the sensory module generate $M_y$ length-$n$
codewords by sampling with replacement from a uniform distribution
on $\T^\d_V$. Assign each codeword a unique index $j\in \My,$
where $\My=\{1,2,\ldots,M_y\}.$ Denote the resulting codebook
\begin{equation*}
        \B_v = \{\v(1),\v(2),\ldots, \v(M_y)\},
\end{equation*}
where the $\v(j)$ are the indexed codewords.

Provide copies of both codebooks $\B_u$ and $\B_v$ to the
recognition module.

\textbf{Memory encoding rule $f$.} Let $\Cx=\{
(\bX(1),1),(\bX(2),2),\ldots,(\bX(M_c),M_c) \}$ be the set of
labeled random patterns to be encoded into memory during the
training phase. We define the memory encoder $f$ in terms of the
following procedure. Given a labeled pattern $(\x(w),w)$, the
memory module searches through the memory codebook $\B_u$ for a
codeword $\u$ such that $(\x(w),\u)\in\T^\d_{XU}$. If such a
codeword is found we denote it by $\u(w)$, and denote its index in
the codebook $\B_u$ by $m(w)$. If $\B_u$ has no codeword that is
strongly jointly $\d$-typical with $\x(w)$, an error is declared
and the label $w$ is associated with the first codeword of $\B_u$.
Denoting the event that the above procedure fails by $E_{1}$ and
its complement by $E^c_{1}$, let
\begin{equation*}
f(\x(w),w) = \left\{%
\begin{array}{ll}
    (1,w) & \hbox{if $E_{1}$ occurs;} \\
    (m(w),w), & \hbox{if $E^c_{1}$ occurs.} \\
\end{array}%
\right.
\end{equation*}
An error is also declared if the above procedure results in
assigning more than one pattern label to the same memory codeword;
denote this second error event $E_{2}$. The training phase
corresponds formally to applying $f$ to all $M_c$ patterns in
$\Cx$, inducing the set
\begin{equation*}
\Cu=f(\C_x)=\{ (m(1),1),\ldots,(m(M_c),M_c)\}.
\end{equation*}
Note that not all of the codewords in $\B_u$ have been used in the
encoding procedure. Likewise, in the decoding algorithm described
below, we need only consider the subset of codewords $\u\in\B_u$
whose indices in $\B_u$ also appear in $\Cu$. We denote the set of
indices for these `active' codewords
$\L=\L(\Cu)=\{m(1),m(2),\ldots,m(M_c)\}$.

After training, reveal the memory codebook $\B_u,$ the compressed
data $\Cu$, and the mapping $f$ to the recognition module.

\textbf{Sensory encoding rule $\phi$.} The sensory encoding rule
$\phi$ is defined as follows. Let $\y$ be an input to the sensory
module during the testing phase. The sensory module searches
sequentially through the sensory codebook $\B_v$ for a codeword
$\v$ such that $\y\v\in\T^\d_{YV}$. If the search succeeds, denote
the found codeword by $\v(\y)$ and denote its index by $\mu(\y)$.
If the search fails, declare an error, and let the sensory encoder
output be $\mu=1$. Letting $E_3$ be the error event and $E^c_3$
its complement, let
\begin{equation*}
\phi(\y) = \left\{%
\begin{array}{ll}
    1, & \hbox{if $E_3$ occurs;} \\
    \mu(\y), & \hbox{if $E^c_3$ occurs.} \\
\end{array}%
\right.
\end{equation*}
Reveal the sensory codebook $\B_v$ and the mapping $\phi$ to the
recognition module.

\textbf{Classifier: $g_1$.} We next specify $g_1$, the first part
of the classifier $g= g_1\circ g_2$. Upon receiving the index
$\mu=\mu(\y)$ from the sensory module, the recognition module
retrieves the $\mu$-th codeword $\v(\y)$ from the sensory codebook
$\B_v$, then searches the `active' portion of the memory codebook
$\B_u(\L)\subset\B_u$ for a codeword $\u$ such that
$\u\v(\y)\in\T^\d_{UV}$. If such a $\u$ exists and is unique,
denote it by $\hat{\u}=\hat{\u}(\mu)$ and its index in the
codebook $\B_u$ by $\mh=\mh(\hat{\u})$. If no such $\u$ exists,
declare an error, $E_4$; if more than one such $\u$ exists,
declare an error $E_5$; and in case of either $E_4$ or $E_5$ let
$\hat{m}=1$. Thus, set
\begin{equation*}
g_1(\mu)=\left\{%
\begin{array}{ll}
    1, & \hbox{if either $E_4$ or $E_5$ or both occur;} \\
    \mh, & \hbox{if both $E^c_4$ and $E^c_5$ occur.} \\
\end{array}%
\right.
\end{equation*}

\textbf{Classifier: $g_2$.} After determining the codeword index
$\mh=g_1(\mu)$, the recognition module searches the set of stored
data $\Cu$ for a pair $(m,w)$ whose first entry is $m=\mh$ and
retrieves the associated class label. Note that if none of the
errors $E_i$, $i=1,\dots,5$ occurs, there pair $(\hat{m},\hat{w})$
is in fact unique. If there is more than one such pair, then to
ensure uniqueness choose the first. Denoting the retrieved label
by $\wh$, let
\begin{equation*}
g_2(\mh,\Cu) = \wh.
\end{equation*}

\textbf{Analysis of the probability of error}

We now show that the probability of recognition error using the
code $(f,\phi,g)_n$ developed above vanishes as $n\ra \infty$. The
following list qualitatively describes all possible sources of
error using the code $(f,\phi,g)_n$:

\begin{description}
    \item[$E_0$] The sensory data is too ambiguous-- i.e. it is not strongly jointly typical with the training pattern;
    \item[$E_1$] The training pattern is unencodable;
    \item[$E_2$] Two or more training patterns are associated with same memory
        codeword;
    \item[$E_3$] The sensory data is unencodable;
    \item[$E_4$] The encoded sensory data matches no codeword in memory;
    \item[$E_5$] The encoded sensory data matches one or more incorrect
        memory codewords.
\end{description}
More formally, the possible errors are
\begin{eqnarray*}
    E_0 &=& \{ (\x(w),\y)\notin \T^\d_{XY} \} \\
    E_1 &=& E_0^c \cap \left\{ \bigcap_{i=1}^{M_x} \left\{ (\x,\u(i))\notin \T^\d_{XU} \right\} \right\} \\
    E_2 &=& \left(\bigcap_{n=0}^1 E_n^c \right)\cap \left\{ \bigcup_{\x(w')\in \mathcal{\C}_x,\;w'\neq w}
    \left\{ (\x(w'),\u(w))\in \T^\d_{XU} \right\}  \right\} \\
    E_3 &=& E_0^c \cap \left\{ \bigcap_{i=1}^{M_y} \left\{ (\y,\v(i))\notin \T^\d_{YV} \right\} \right\} \\
    E_4 &=& \left(\bigcap_{n=0}^3 E_n^c \right)\cap \left\{ (\x(w),\u(w),\y,\v(\y))\notin
    \T^\d_{UXYV},\right\} \\
    E_5 &=& \left(\bigcap_{n=0}^4 E_n^c \right)\cap \left\{ \bigcup_{\u(m')\in \B_u,\;m'\in\L^*}
    \left\{ (\u(m'),\v(\y))\in \T^\d_{UV} \right\}  \right\}, \\
\end{eqnarray*}
where in the last line the set $\L^*$ includes all indices in $\L$
except $m(w)$, i.e. $\L^*=\L \setminus m(w)$. The average total
probability of error is upper-bounded as
\begin{eqnarray*}
P_e^n \leq Pr\left\{\bigcup_{\ell=1}^{5} \right\}\leq
\sum_{\ell=0}^{5}P(E_\ell).
\end{eqnarray*}
Hence to show $P_e^n\leq \eps$ it suffices to show that each term
in the sum vanishes as $n\ra \infty$.

\def\X {{ \mathbf{X} }}
\def\Y {{ \mathbf{Y} }}
\def\U {{ \mathbf{U} }}
\def\V {{ \mathbf{V} }}

\emph{Encoding Errors}

\emph{Error event $E_0$}: By the Asymptotic Equipartition
Property, $Pr(E_0)\ra 0$ \cite{cover-thomas-book}.

\emph{Error event $E_1$}: For $E_1$, we use the well known fact
that if $R_x\geq I(X;U)$, then the $M_x=2^{nR_x}$ codewords in
$\B_u$ are sufficient to cover the pattern source
\emph{\textbf{X}}. Explicitly, let $R_x=I(X;U)+\alpha$, for any
$\alpha>0$. Then for any $\eps>0$ and sufficiently large $n$,
\begin{eqnarray*}
        Pr(E_1) &=& \sum_{\x\y\in\T^\d} Pr(E_1|\x)Pr(\x)Pr(\y|\x) \\
        &\leq& \sum_{\x\in\T^\d} Pr(E_1|\x)Pr(\x) \\
        &=& \sum_{\x\in\T^\d} \{1-Pr(\x\bU\in\T^\d|\x)\}^{M_x}Pr(\x)\\
        &\stackrel{a}{\leq}& \{1-2^{-n[I(X;U)+\alpha/2]}\}^{M_x} \\
        &\stackrel{b}{\leq}& 2^{-M_x 2^{-n[I(X;U)+\alpha/2]}} \\
        &\leq& 2^{- 2^{n[I(X;U)+\alpha-I(X;U)-\alpha/2]}} \\
        &=& 2^{-2^{n\alpha/2}} \\
        &\leq& \eps,
\end{eqnarray*}
where (a) is due to the property of strongly jointly typical sets
in equation \ref{Eqn: TypicalSetProp}, and in (b) we have used
$(1-\alpha)^\beta \leq 2^{-\alpha\beta}$. Hence, $Pr(E_1)\ra 0$.

\emph{Error event $E_2$:} Conditioned on $E_0^c\cap E_1^c$, we
have $\u(w)\in \T^\d_U$. The sequences $\X(w')\in\Cx,\; w' \neq w$
are generated
    independently of $\u(w)$. Thus
\begin{eqnarray*}
            P(E_2) &=& \sum_{\bX(w')\in \Cx,w'\neq w}
            Pr\left( \tilde{\X}\u(w)\in \T^\d_{XU} |\u(w)\in \T^\d_U \right) \\
            &\leq& |\Cx| 2^{-nI(X;U)+n\eps} \\
            &\leq& 2^{nR_c}2^{-nI(X;U)+n\eps} \\
            &\leq& \eps
\end{eqnarray*}
for large enough $n$, since $R_c\leq I(U;V) \leq I(X;U)$ under the
Markov assumption $U-X-Y-V$. Hence, $P(E_2)\ra 0$.

\emph{Error event $E_3$:} By a covering argument analogous to the
one used in the analysis of event $E_1$, having $M_y\geq
2^{nI(Y;V)}$ codewords in $\Cv$ is sufficient to ensure $P(E_3)
\ra 0$.

\emph{Decoding errors}

\def\ba {{ \mathbf{a} }}
\def\bb {{ \mathbf{b} }}
\def\bc {{ \mathbf{c} }}
\emph{Error event $E_4$:} To analyze the probability of event
$E_4$, we invoke the following uniform version of the well-known
Markov Lemma
\cite{berger77MultTermSC,Tung78Thesis,kaspi79Thesis,kaspiBeger82}.
\begin{lemma}
Let $A-B-C$ be a Markov chain; let $\ba\bb\in\T^\d_{AB}$; let
$\bC$ be chosen from a uniform distribution over $\T^\d_{bC}$; and
let $\eps>0$ be any positive constant. Then $Pr(\ba\bb\bC\notin
\T^\d_{ABC})\leq \eps$ for $n$ sufficiently large.
\end{lemma}

To bound the probability of event $E_4$, we condition on
$\cap_{i=0}^3 E_i^c$ and apply the Markov lemma twice in
succession to establish the following two claims:
\begin{eqnarray*}
\text{i)} && Pr(\x\y\bV(\y) \notin \T^\d_{XYV}|\x\y\in\T^\d,
\bV(\y)\in\T^\d_{yV})\leq
\eps \\
\text{ii)} && Pr(\bU(w)\x\y\v(\y) \notin \T^\d_{XYUV}|
\x\y\v(\y)\in\T^\d,\bU(w)\in\T^\d_{\x\U})\leq \eps
\end{eqnarray*}

To prove (i), note that the conditions of the Markov Lemma can be
satisfied making the following substitutions in the Lemma:
$(\ba,\bb,\bC) \ra (\x,\y,\V(\y))$. Similarly, to prove (ii), put
$(\ba,\bb,\bC)\ra (\y\v,\x,\U(w))$. Combining (i) and (ii), we
conclude that $\Pr(E_4)\ra 0$.

\emph{Error event $E_5$:}

Given $\bigcap_{n=0}^3 E_n^c$, we  have $\v(\y)\in \T^\d_V$. The
sequences $\bU(m')\in\B_u$, $m'\in \L^*=\L \setminus m(w)$ were
generated independently of $\v(\y)$. Thus
\begin{eqnarray*}
            P(E_5) &=& \sum_{\bU(m') \in \B_u, m'\in\L^*}
            Pr\left(  \bU(w')\v(\y)\in T_{UV} |\v(\y)\in T_V \right) \\
            &\leq& |\L^*| 2^{-n[I(U;V)-\eps]} \\
            &\leq& 2^{nR_c}2^{-n[I(U;V)-\eps]} \\
            &\leq& \eps
\end{eqnarray*}
for large enough $n$, since $R_c\leq I(U;V)$. Hence, $P(E_5) \ra
0$.

We have constructed a rate $\bR$ code for which $P_e^n \leq
\sum_{n=0}^5 Pr\{ E_n\}\ra 0.$ Consequently, $\bR\in\mathcal{R}$,
completing the proof.

%%%%%%%%%%%%%%%%%%%%%%%%%%%%%%
\end{proof}

\section{Proof of the outer bound}\label{Sec: Outer bound proof}
In this section we prove theorem \ref{Thm: outer bound}, which
states the outer bound $\R\subseteq \Ro$. In the proof let $W$ be
the test index, selected from a uniform distribution $p(w)$ over
the pattern indices $\Mc$; let $\bX=\bX(W)$ be the selected test
pattern from the set of training patterns $\Cx$; let $m=m(W)$ be
the compressed, memorized form of $\bX$ computed from $f$ as
$(m,W)=f(\bX,W)$; let $\Cu=f(\Cx)$ be the memorized data; let
$\bY$ be the sensory data; and let $\mu=\mu(W)=\phi(\bY)$ be the
encoded form of sensory data. Note that $m$ and $\mu$ are random
variables through their dependence on $\bX$ and $\bY$. The mutual
informations in the proof are calculated with respect to the joint
distribution over $(W,\Cx,\Cu,\bX,\bY,m,\mu,\hat{w})$. We can
verify that this distribution is well-defined by writing it out
explicitly:
\begin{equation*}
    p(w,\Cx,\Cu,\x,\y,m,\mu)=p(w)p(\Cx)p(\Cu|\Cx)p(\x|w,\Cx)p(\y|\x)p(m|\x,w)p(\mu|\y),
\end{equation*}
where
\begin{eqnarray*}
    p(w) &=&  \begin{cases}
            \frac{1}{M_c} & w\in\mathcal{M}_c, \\
            0 & \text{otherwise};
            \end{cases} \\
    p(\Cx) &=& \prod_{i=1}^{M_c}\prod_{i=1}^{n}p(x_i) \\
    p(\Cu|\Cx) &=& \begin{cases}
            1 & \Cu=f(\Cx) \\
            0 & \text{otherwise};
        \end{cases} \\
    p(\x|w,\Cx) &=& \begin{cases}
            1 & \x=\x(w), (\x,w)\in\Cx \\
            0 & \text{otherwise};
        \end{cases} \\
    p(\y|\x) & = & \prod_{i=1}^n p(y_i|x_i) \\
    p(m|\x,w) &=&
        \begin{cases}
            1 & f(\x,w)=(m,w) \\
            0 & \text{otherwise};
        \end{cases} \\
       p(\mu|\y) & = &
        \begin{cases}
            1 & \mu = \phi(\y) \\
            0 & \text{otherwise.}
        \end{cases} \\
        p(\hat{w} | \mu,\Cu) &=&
        \begin{cases}
            1 & \hat{w} = g(\mu,\Cu) \\
            0 & \text{otherwise.}
        \end{cases}
\end{eqnarray*}
The independence relationships underlying the structure of this
distribution are clear from the block diagram of figure \ref{Fig:
MainDiagram}. They are also usefully displayed using a directed
graphical model (`Bayes' net') \cite{freyBook,jordanGMintro}.

%\input{Fig_IndependenceRelns}
%\graphicspath{{MatlabFigures/IllustratorPics/}}
\begin{figure}[htbp]
\begin{center}
    \psfrag{Bx}{$X^n$} \psfrag{Y}{$Y^n$} \psfrag{m}{$m$}
    \psfrag{mu}{$\mu$} \psfrag{Cx}{$\Cx$} \psfrag{Cu}{$\Cu$}
    \psfrag{\wh}{$\hat{w}$} \psfrag{W}{$W$}
    \includegraphics[width=2.5in]{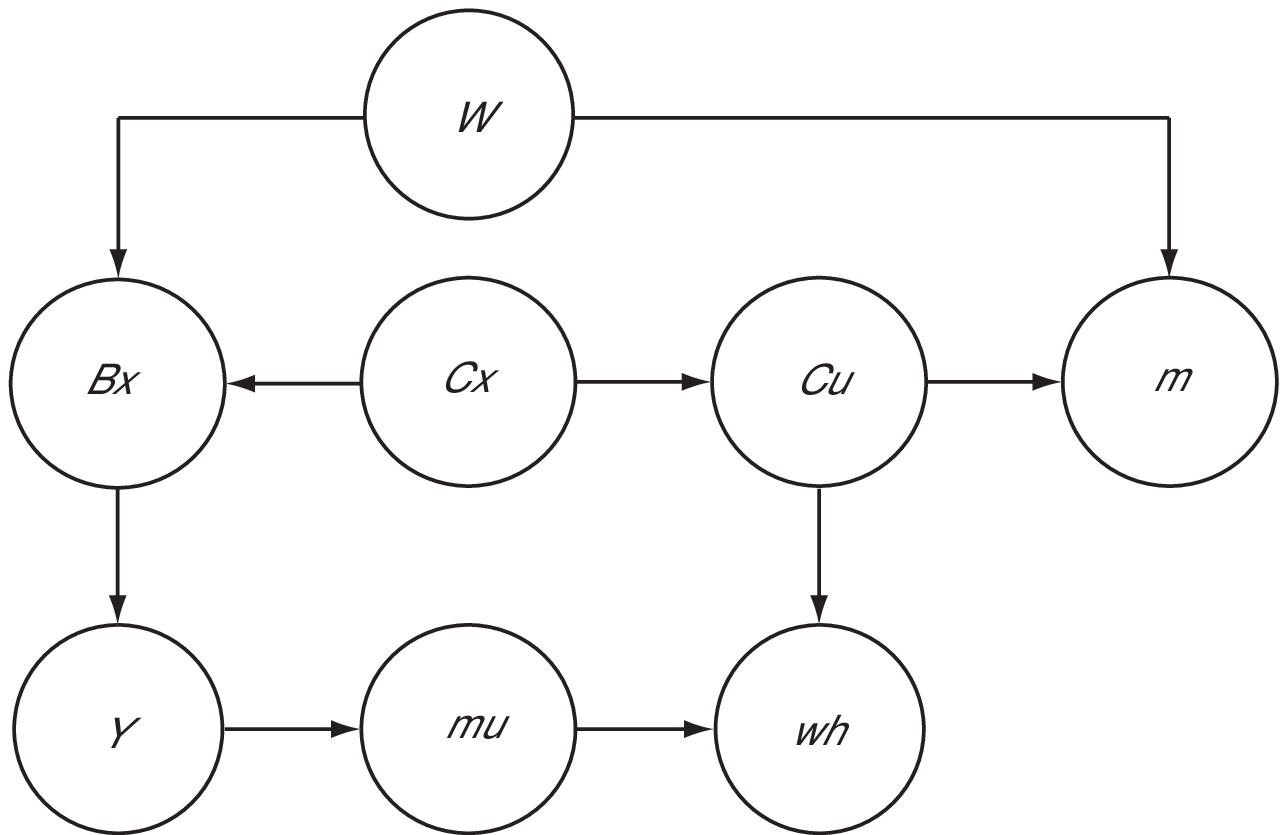}
    \caption{Independence relationships for $(W,\Cx,\Cu,\bX,\bY,m,\mu,\hat{w})$}
    \label{Fig:GaussSurfaces}
\end{center}
\end{figure}

\begin{proof}(Theorem \ref{Thm: outer bound})

Assume $\bR=(R_x,R_y,R_c)\in\R$. Then there exists a sequence of
$(M_x,M_y,M_c,n)$ codes $(f,\phi,g)_n$, such that for any $\eps
>0,$
\begin{eqnarray*}
        M_c & \geq & 2^{nR_c} \\
        M_x & \leq & 2^{nR_x} \\
        M_y & \leq & 2^{nR_y}
\end{eqnarray*}
and $P_e^n=Pr(\hat{W}\neq W) \leq \eps.$ To show that $\bR\in\Ro$,
we must construct a pair of auxiliary random variables $UV$ such
that $UV\in\Po$ and $\bR\in\R_{UV}$.

We construct the desired pair $UV$ in three steps: (1) We
introduce a set of intermediate random variable pairs $U_i V_i,
i=1,2,\ldots,n$, individually contained in $\Po$; (2) we derive
mutual information inequalities for $R_x$, $R_y$, and $R_c$
involving sums of the intermediate variables; (3) we convert the
sum inequalities into inequalities in the final pair $UV$ by
applying Lemma \ref{Lemma: sums to single letter}.

\emph{Step 1:}

Let the intermediate auxiliary random variables be
\begin{eqnarray*}
U_i&=&(m,W,X^{i-1}) \\
V_i&=&(\mu,Y^{i-1}),
\end{eqnarray*}
for $i=1,2,\ldots,n$. Each pair is in $\Po$. This is verified for
the $U_i$ by calculating
\begin{eqnarray*}
I(U_i;Y_i|X_i) &=& H(Y_i|X_i)-H(Y_i|m,W,X^{i-1}, X_i) \\
&=& H(Y_i|X_i)-H(Y_i|m,W,X^i) \\
&\stackrel{a}{\leq}& H(Y_i|X_i)-H(Y_i|m,W,X^n) \\
&\stackrel{b}{=}& H(Y_i|X_i)-H(Y_i|X^n) \\
&\stackrel{c}{=}& H(Y_i|X_i)-H(Y_i|X_i) \\
&=& 0,
\end{eqnarray*}
where the reasons for the lettered steps are (a) conditioning
reduces entropy, (b) the $Y_i$ are independent of all other
variables given $X^n$, and (c) the pairs $X_i Y_i$ are i.i.d.
Hence, $U_i-X_i-Y_i$ is a Markov chain. By a similar argument,
$X_i-Y_i-V_i$ is also a Markov chain. Hence, $U_i V_i\in\Po$ for
each $i=1,2,\ldots,n$.

\emph{Step 2:}

First,
\begin{eqnarray*}
    M_c(nR_x) &\geq& M_c\log M_x \\
    & \geq & H(\Cu) \\
    &\stackrel{a}{=}& H(\Cu)-H(\Cu|\Cx) \\
    &=& I(\Cu;\Cx) \\
    &=& H(\Cx)-H(\Cx|\Cu) \\
    &\stackrel{b}{=}& \sum_{w=1}^{M_c} [H(X^n(w),w)-H(X^n(w),w|m(w),w)] \\
    &\stackrel{c}{=}& \sum_{w=1}^{M_c} [H(X^n(W)|W=w)-H(X^n(W)|m(w),W=w)] \\
    &\stackrel{d}{=}& \sum_{w=1}^{M_c} [H(X^n(w))-H(X^n(W)|m(w),W=w)] \\
    &\stackrel{e}{=}& \sum_{w=1}^{M_c} [H(X^n)-H(X^n|m,W=w)] \\
    &\stackrel{f}{=}& \sum_{w=1}^{M_c}\sum_{i=1}^n [H(X_i)-H(X_i|m,W=w,X^{i-1})] \\
    &\stackrel{g}{=}& M_c \sum_{i=1}^{n} \sum_{w=1}^{M_c} [H(X_i)-H(X_i|m,W=w,X^{i-1})]p(w) \\
    &=& M_c \sum_{i=1}^{n} [H(X_i)-\sum_{w=1}^{M_c} p(w) H(X_i|m,W=w,X^{i-1})] \\
    &=& M_c \sum_{i=1}^{n} [H(X_i)- H(X_i|m,W,X^{i-1})]\\
    &\stackrel{h}{=}& M_c \sum_{i=1}^n [H(X_i)-H(X_i|U_i)] \\
    &=& M_c \sum_{i=1}^n I(X_i;U_i),
\end{eqnarray*}
or
\begin{equation*}
nR_x \geq \sum_{i=1}^n I(X_i;U_i),
\end{equation*}
where the justifications are (a) $\Cu=f(\Cx)$; (b) the pairs
$(X^n(w),w)$ are independent; (c) in this expression $w$ is a
deterministic variable (i.e. $H(w)=H(W|W=w)=0$); (d) the $X^n(w)$
are i.i.d. and independent of their index $w$; (e) to simplify
notation, we have written $m=m(w)$, $X^n=X^n(w)$; (f) the $X_i$
are i.i.d.; and (g) $W$ is distributed according to $p(w)=1/M_c$,
$w=1,2,\ldots,M_c$.

Next,

\begin{eqnarray*}
    nR_y &\geq& H(\mu) \\
    &\stackrel{a}{=}& H(\mu)-H(\mu|Y^n) \\
    &=& \sum_{i=1}^n H(Y_i)-H(Y_i|Y^{i-1}\mu) \\
    &=& \sum_{i=1}^n H(Y_i)-H(Y_i|V_i) \\
    &=& \sum_{i=1}^n I(Y_i;V_i). \\
\end{eqnarray*}
Step (a) follows from $\mu=\phi(Y^n)$.

Finally,

\def\m {{ \mu }}
\def\e {{ \epsilon_n }}
\begin{eqnarray*}
    nR_c&\leq& \log M_c \\
    &=& H(W) \\
    &=& I(W;\Cu,\mu)+H(W|\Cu,\mu) \\
    &\stackrel{a}{\leq}& I(W;\Cu,\mu) + n\e \\
    &=& I(W;\Cu)+I(W;\mu|\Cu) +n\e \\
    &\stackrel{b}{=}& 0+I(W;\mu|\Cu) +n\e \\
    &=& I(W,\Cu;\mu)-I(\mu;\Cu) +n\e \\
    &\leq& I(W,\Cu;\mu)+n\e \\
    &\stackrel{c}{=}& I(W,m;\mu)+ n\e \\
    & \stackrel{d}{=} & \sum_{i=1}^n I(X_i;U_i)+I(Y_i;V_i)-I(X_i Y_i;U_i V_i) + 2n\e\\
    & \stackrel{e}{=}& \sum_{i=1}^n I(U_i;V_i)-I(U_i;V_i|X_i V_i)
    +2n\e,
 \end{eqnarray*}
The lettered steps are justified as follows.

(a) By assumption, $Pr(\hat{w}\neq W)=P_e^n \rightarrow 0$, where
$\hat{w}=g_n(\mu,\Cu)$. Thus, applying Fano's inequality yields
      \begin{equation*}
        H(W|\Cu,\mu) \leq H(P_e^n)+P_e^n\log(M_c-1)\leq n\epsilon_n,
      \end{equation*}
      where $\eps_n\ra 0$.

(b)The test index $W$ and patterns $\Cx$ are drawn independently,
hence $W$ and $\Cu=f(\Cx)$ are independent and $I(W;\Cu)=0$.

(c) Writing $\Cu=\Cu^*\cup \{(m,W)\}$ , $\Cu^*=\Cu \setminus
\{(m,W) \}$, we have
\begin{eqnarray*}
    I(W,\Cu;\mu)&=& I(W,(m,W),\Cu^*;\mu) \\
    &=& I(W,m;\mu)+I(W,\Cu^*;\mu|W,m) \\
    &=& I(W,m;\mu)+I(\Cu^*;\mu|W,m) \\
    &=& I(W,m;\mu)+0,
\end{eqnarray*}
since the $(m(i),i)$ are independent of $\mu$ for $i\neq W$.

(d) To justify this step we invoke the following two results,
proved in Appendix \ref{Appdx: mixing Lemmas}. Let
      $A,\a,B,\b,$ and $\g$ be arbitrary discrete random
      variables. Then:
      \begin{theorem}\label{Thm: ab lemma}
      \begin{equation*}
            I(\a;\b) \geq I(A;a)+I(B;\b)-I(AB;\a\b),
      \end{equation*}
      with equality if and only if $I(A\a;B\b)=I(A;B)$.
      \end{theorem}

      \begin{theorem}\label{Thm: Gelfand Pinsker lemma}
      Let $Z_i=(\gamma;A^{i-1}),\;i=1,2,\ldots,n$, where the $A_i$ are i.i.d. Then
      \begin{equation*}
      \sum_{i=1}^n I(A_i;Z_i) = I(A^n;\g).
      \end{equation*}
      \end{theorem}

    To apply Theorem \ref{Thm: ab lemma}, make the substitution
      $(\a,\b,A,B)\ra (mW,\mu,X^n,Y^n)$. Then the condition for
      equality is satisfied:
      \begin{eqnarray*}
        I(X^n,m,W;Y^n,\mu) &=& I(X^n,W;Y^n,\mu) +
        I(m,W;Y^n,\mu|X^n,W) \\
        &\stackrel{a}{=}& I(X^n,W;Y^n,\mu)+0 \\
        &=& I(X^n,W;Y^n)+I(X^n,W;\mu|Y^n) \\
        &\stackrel{b}{=}& I(X^n,W;Y^n)+0 \\
        &=& I(X^n;Y^n)+I(W;Y^n|X^n) \\
        &\stackrel{c}{=} & I(X^n;Y^n)+0,
      \end{eqnarray*}
    since (a) $(m,W)=f(X^n,W)$, (b) $\mu=\phi(Y^n)$, and (c) $Y^n$ only depends on $W$
    through $X^n=X^n(W)$, so that $H(Y^n|X^n,W)=H(Y^n|X^n)$. Thus Theorem \ref{Thm: ab
    lemma} yields
        \begin{equation}\label{Eqn: first lemma result}
            I(m,W;\m) = I(X^n;m,W)+I(Y^n;\m)-I(X^n,Y^n;m,W,\m).
      \end{equation}

    Next, apply Theorem \ref{Thm: Gelfand Pinsker lemma} three
    times with the substitutions:
    \begin{eqnarray*}
    (Z_i,\g,A^{i-1}) &\ra& (U_i,mW,X^{i-1}) \\
    &\ra& (V_i,\mu,Y^{i-1}) \\
    &\ra& (U_i V_i, m W \mu, X^{i-1} Y^{i-1}),
    \end{eqnarray*}
    to obtain
    \def\x {{ X_i }}
    \def\X {{ X^{i-1} }}
    \def\y {{ Y_i }}
    \def\Y {{ Y^{i-1} }}
    \def\u {{ U_i }}
    \def\v {{ V_i }}
    \begin{eqnarray*}
    \sum_{i=1}^n I(X_i;U_i) &=& I(X^n;m,W) \\
    \sum_{i=1}^n I(Y_i;V_i) &=& I(Y^n;\mu) \\
    \sum_{i=1}^n I(X_i Y_i;U_i V_i) &=& I(X^n Y^n;m,W,\mu) \\
    \end{eqnarray*}

    Adding the first two expressions and subtracting the third
    yields

       \begin{equation}\label{Eqn: second lemma result}
            \sum_{i=1}^n [I(\x;\u)+I(\y;\v)-I(\x,\y;\u,\v)]
            = [I(X^n;m,W)+I(Y^n,\m)-I(X^n,Y^n;m,W,\m)].
        \end{equation}

        Combining (\ref{Eqn: first lemma result})
        and (\ref{Eqn: second lemma result}) yields
        \begin{equation*}
            I(m,W;\mu)= \sum_{i=1}^n I(X_i;U_i)+I(Y_i;V_i)-I(X_i,Y_i;U_i,V_i),
        \end{equation*}
        as claimed.

(d) This step is justified by the following chain of equalities:
  \begin{equation*}
    \begin{split}
    I(X_i;U_i)+I(Y_i;V_i)&{}-I(X_i,Y_i;U_i,V_i) \\
    &{}=H(U_i)-H(U_i|X_i)+H(Y_i)-H(V_i|Y_i)-[H(U_i V_i)-H(U_i V_i|X_i Y_i) \\
    &{}=[H(U_i)+H(V_i)-H(U_i V_i)]-[H(U_i|X_i)+H(V_i|Y_i)-H(U_i V_i|X_i Y_i)] \\
    &{}=I(U_i;V_i) - [H(U_i|X_i Y_i)+H(V_i|X_i Y_i)-H(U_i V_i|X_i Y_i)] \\
    &{}=I(U_i;V_i)-I(U_i;V_i|X_i Y_i),
    \end{split}
  \end{equation*}
  for each $i=1,2,\ldots,n$, where in the second-to-last step we have used the
  fact that $U_i-X_i-Y_i$ and
  $X_i-Y_i-V_i$ are Markov chains for $i=1,2,\ldots,n$, as shown
  above in Step 1.

\emph{Step 3:}

For this step we use the following Lemma, proved in Appendix
\ref{Appdx: Convexity of Rout}:
\begin{lemma}\label{Lemma: sums to single letter}
Suppose $U_i V_i\in \Po, i=1,2,\ldots,n$. Then there exists
$UV\in\Po$ such that
\begin{eqnarray*}
\frac{1}{n}\sum_{i=1}^n I(X_i;U_i) &=& I(X;U) \\
\frac{1}{n}\sum_{i=1}^n I(Y_i;V_i) &=& I(Y;V) \\
\frac{1}{n}\sum_{i=1}^n I(U_i;V_i)-I(U_i;V_i|X_iY_i)&=& I(U;V)-I(U;V|XY) \\
\end{eqnarray*}
\end{lemma}

Applying Lemma \ref{Lemma: sums to single letter} to the results
of steps 1 and 2, we obtain
\begin{eqnarray*}
R_x &\geq& \frac{1}{n} \sum_{i=1}^n I(X_i;U_i) = I(X;U) \\
R_y &\geq& \frac{1}{n} \sum_{i=1}^n I(Y_i;V_i) = I(Y;V) \\
R_c &\leq& \frac{1}{n} \sum_{i=1}^n I(U_i;V_i)-I(U_i; V_i|X_i Y_i) \\
&=& I(U;V)-I(U;V|XY)
\end{eqnarray*}
where $UV\in\Po$. With respect to this $UV$, by definition we have
$\bR\in\R_{UV}$. Hence, $\bR\in\Ro$, and the proof is complete.

\end{proof}

\section{Convexity of the outer bound}\label{Appdx: Convexity of Rout}

In this Appendix we prove a slightly more general version of Lemma
\ref{Lemma: sums to single letter} from section \ref{Sec: Outer
bound proof}, and demonstrate that the outer bound rate region
$\Ro$ is convex.

In the following, let $\Q$ be any finite alphabet, and assume that
we have pairs $X_q Y_q$ for all $q\in\Q$ which are i.i.d. $\sim
p(xy)$.

\begin{lemma}\label{Lemma: sums to single letter, general version}
Suppose $U_q V_q\in\Po$ for all $q\in\Q$, and let let $Q\sim p(q),
q\in\Q$ be any discrete random variable independent of the pairs
$\{X_q Y_q\}$. Then there exists a pair of discrete random
variables $UV\in\Po$ such that
\begin{eqnarray*}
\sum_{q\in\Q} p(q)I(X_q;U_q) &=& I(X;U) \\
\sum_{q\in\Q} p(q)I(Y_q;V_q) &=& I(Y;V) \\
\sum_{q\in\Q} p(q)[I(U_q;V_q)-I(U_q;V_q|X_q Y_q)&=&
I(U;V)-I(U;V|XY).
\end{eqnarray*}
\end{lemma}

\begin{remark}
Lemma \ref{Lemma: sums to single letter} in section \ref{Sec:
Outer bound proof} follows immediately from the above Lemma, by
choosing $\Q=\{1,2,\ldots,n\}$ and $p(q)=1/n$ for all $q\in\Q$.
\end{remark}

\begin{proof}
As a candidate for the pair $UV$ in the Lemma, consider
$U=(U_Q,Q)$ and $V=(V_Q,Q)$, i.e.
\begin{eqnarray*}
U &=& \{U_q \text{ if } Q=q \} \\
V &=& \{V_q \text{ if } Q=q \}.
\end{eqnarray*}
To verify that $UV\in\Po$, we proceed to check that $U-X-Y$ and
$X-Y-V$ are Markov chains.

By the assumption $U_qV_q\in\Po$ for each $q\in\Q$, we have
$I(U_q;Y_q|X_q)=0$ and $I(V_q;X_q|Y_q)=0$. Hence
\begin{eqnarray*}
0 &=& \sum_{q\in\Q} p(q)I(U_q;Y_q|X_q) \\
&=& \sum_{q\in\Q} p(q)I(U_q;Y_q|X_q, Q=q) \\
&=& I(U_Q;Y_Q|X_Q Q) \\
&\stackrel{a}{=}& I(U_Q;Y|X,Q) \\
&=& I(U_Q Q;Y|X)-I(Q;Y|X) \\
&\stackrel{b}{=}& I(U_Q Q;Y|X) \\
&=& I(U;Y|X),
\end{eqnarray*}
where in (a) we are able to drop the subscript $Q$ on $X_Q$ and
$Y_Q$ because the $X_q$ and $Y_q$ are i.i.d. and independent of
$Q$; and similarly (b) is because $I(Q;Y|X)=0$, due to the
independence of $Q$ and $Y$. By an analogous calculation, we also
find $I(V;X|Y)=0$. Hence, $U-X-Y$ and $X-Y-V$, and $UV\in\Po$ as
desired.

It remains to demonstrate the three equalities in the Lemma. For
the first equality, we write
\begin{eqnarray*}
    I(X;U) &=& I(X;U_Q Q) \\
    &=& I(X;U_Q|Q)+I(X;Q) \\
    &\stackrel{a}{=}&  I(X;U_Q|Q) \\
    &\stackrel{b}{=}& I(X_Q;U_Q|Q) \\
    &=& \sum_{q\in\Q} p(q)I(X_q;U_q),
\end{eqnarray*}
where, as above, (a) and (b) follow from the facts that the $X_q$
are i.i.d. and independent of $Q$. Similar calculations yield
\begin{equation*}
I(Y;V) = \sum_{q\in\Q} p(q)I(Y_q;V_q),
\end{equation*}
which is the second required equality, and
\begin{equation*}
I(XY;UV) = \sum_{q\in\Q} p(q)I(X_qY_q;U_q V_q).
\end{equation*}
This last equality can be combined with the first two to yield the
third required equality using
\begin{equation*}
I(X;U)+I(Y;V)-I(XY;UV)=I(U;V)-I(U;V|XY).
\end{equation*}
which follows from the two short Markov chains $U-X-Y$ and
$X-Y-V$, as shown in subsection \ref{subsec: surface expressions},
equation \ref{eqn: alt form}. The proof is complete.
\end{proof}

The convexity of $\Ro$ follows readily from the preceding Lemma.
\begin{lemma}\label{Lemma: convexity of Rout}
$\Ro$ is convex. That is, let $R_q$ be any set of rates such that
$R_q\in\Ro$ for all $q\in\Q$, where $\Q$ is a finite alphabet, and
let $p(q)$ be any probability distribution over $\Q$. Then
$R=\sum_{q\in\Q} p(q) R_q \in \Ro$.
\end{lemma}

\begin{proof}
Fix an arbitrary distribution $p(q)$ and rates $\bR_q\in\Ro$ for
all $q\in\Q$. By the definition of $\Ro$, for each rate $\bR_q$,
there exists a pair $U_q V_q\in\Po$ such that $\bR_q\in\R_{U_q
V_q}$. Consequently,
\begin{eqnarray*}
R_x = \sum_{q\in\Q} p(q)R_{x,q} &\geq&
\sum_{q\in\Q}p(q)I(X_q;U_q)\\
R_y = \sum_{q\in\Q} p(q)R_{y,q} &\geq&
\sum_{q\in\Q}p(q)I(Y_q;V_q)\\
R_c = \sum_{q\in\Q} p(q)R_{c,q} &\leq& \sum_{q\in\Q}p(q)I(U_q;
V_q)-I(U_q;V_q|U_q V_q).
\end{eqnarray*}

As in the proof of Lemma \ref{Lemma: sums to single letter,
general version}, use these pairs to construct a new pair $UV$, by
defining $U=(U_Q,Q)$, $V=(V_Q,Q)$. From the proof of Lemma
\ref{Lemma: sums to single letter, general version}, we know (1)
that $UV\in\Po$, and (2) the sums on the right hand sides of the
inequalities above can be replaced with expressions in $U$ and
$V$, yielding
\begin{eqnarray*}
R_x &\geq& I(X;U)\\
R_y &\geq& I(Y;V)\\
R_c &\leq& I(U; V)-I(U;V|U V),
\end{eqnarray*}
which means that $R \in\R_{UV}$ for the given $UV$. Hence,
$\bR=\sum_{q\in\Q}p(q)\bR_q \in\Ro$. Since $p(q)$ and
$\bR_q\in\Ro$ were arbitrary, we conclude that $\Ro$ is convex.
\end{proof}

\section{Proof of theorem \ref{theorem: extension}}\label{Appendix: proof of extension}
In this section we prove theorem \ref{theorem: extension}. The
argument is based on time sharing.  Consider a sequence of codes
of lengths $n_i$ that achieve $(R_c, R_x, R_y)$. Corresponding to
this sequence is a sequence of codes of lengths $m_i$ that satisfy
$\theta m_i = n_i$, constructed as follows. For each $m_i$, select
any $\theta m_i$ components; reveal the indices of the selected
components to the memory encoder and the sensory encoder.  Use the
corresponding code from the first sequence on these components,
ignoring all other components.  For $m_i$, there are $2^{m_i
\theta R_c}$ patterns, $2^{m_i \theta R_x}$ memory states, and
$2^{m_i \theta R_y}$ sensory states.

The corollary \ref{corollary: extension corollary} follows
immediately from the inner bound, theorem \ref{Thm: inner bound}.

\section{Proof of Lemma \ref{desc: starstar} }\label{appendix: starstar proof} In this Appendix we prove
the `if-then' statement asserted in Lemma \ref{desc: starstar}.

The assumptions of the statement are that (a) $\Pm=\Po$; and (b)
that the achievable rate region $\R$ is convex. We wish to show
that these imply $\R=\Ro$.

From theorem \ref{Thm: outer bound}, we have $\Ro\supseteq \R$. To
prove the Lemma, we must demonstrate the converse, $\Ro\subseteq
\R$.

It suffices to show the boundary points of $\Ro$ are achievable.
Let $\bR=(R_c,R_x,R_y)$ be an arbitrary rate on the boundary of
$\Ro$. Then there exists $UV\in\Po$ such that
$R_c=I(X;U)+I(Y;V)-I(XY;UV)$, $R_x= I(X;U)$ and $R_y=I(Y;V)$. In
turn, assumption (a) $\Pm=\Po$ implies that there exists $Q\sim
p(q), q\in\Q$ independent of $XY$ and pairs $U_q V_q\in \Pi,
q\in\Q$ such that $R_x=I(X;U_Q,Q)$, $R_y=I(Y;V_Q,Q)$, and
$R_c=I(X;U_Q,Q)+I(Y;V_Q,Q)-I(XY;U_Q V_Q, Q)$. Hence, using the
independence of $Q$ from $X$ and $Y$ we have
\begin{eqnarray*}
R_x &=& \sum_{q\in\Q} I(U_q;X)p(q) \\
R_y &=& \sum_{q\in\Q} I(V_q;Y)p(q) \\
R_c &=& =\sum_{q\in\Q} [I(U_q;X)+I(V_q;Y)-I(U_q V_q;XY)]p(q).
\end{eqnarray*}
Next, let $R_{xq}=I(U_q;X)$, $R_{yq}=I(V_q;Y)$,
$R_{cq}=I(X;U_q)+I(Y;V_q)-I(XY;U_q V_q)$, for $q=1,2,\ldots,|\Q|$.
Then, by definition, each rate $R_q=(R_{cq},R_{xq},R_{yq})$ is in
$\Ri$. Since $\Ri\subseteq \R$ by theorem \ref{Thm: inner bound},
$R_q\in\R$ for each $q\in\Q$.

According to the preceding argument, $\bR=(R_c,R_x,R_y)$ is a
convex combination of achievable rates. Consequently, if $\R$ is
convex as assumed, then $\bR\in\R$. Since the rate $\bR$ was an
arbitrary boundary point of $\Ro$, we conclude $\Ro\subseteq \R$,
hence $\R=\Ro$ as desired.

\section{Proofs of theorems \ref{Thm: ab lemma} and \ref{Thm: Gelfand
Pinsker lemma}}\label{Appdx: mixing Lemmas}

Consider the elementary Shannon inequalities, stated in the
following two Lemmas. The variables $A,B,\a,\b,\g,\d$ appearing in
the Lemmas denote arbitrary discrete random variables.
\begin{lemma}\label{Lemma_ANoInd}
    \begin{equation*}
        I(A;\a)=I(A;\a,\g)-I(A,\a;\g)+I(\a;\g).
    \end{equation*}
\end{lemma}
\begin{proof}
    \begin{eqnarray*}
        I(A;\g|\a) &=& I(A;\a,\g)-I(A;\a) \\
        &=& I(A,\a;\g)-I(\g;\a).
    \end{eqnarray*}
\end{proof}

\begin{lemma}\label{Lemma_ABSum}
    \begin{equation*}
        I(A;\a)+I(B;\b)=I(A;B)+I(\a;\b)
        -I(A,\a;B,\b)+I(A,B;\a,\b)
    \end{equation*}
\end{lemma}
\begin{proof}
    \begin{equation*}
    \begin{split}
        I(A,\a;&{}B,\b)-I(A,B;\a,\b) \\
        &{}=\phantom{-} H(A,\a)+H(B,\b)-H(A,B)-H(\a,\b)\\
        &{}= -I(A;\a)-I(B;\b)+I(A;B)+I(\a;\b)
    \end{split}
    \end{equation*}
\end{proof}

Theorems \ref{Thm: ab lemma} and \ref{Thm: Gelfand Pinsker lemma}
follow directly from the Lemmas above.

\begin{theorem}[Theorem \ref{Thm: ab lemma}]\label{Lemma_ABIneq}
    \begin{equation*}
        I(\a;\b)\geq I(A;\a)+I(B;\b)-I(A,B;\a,\b)
    \end{equation*}
    with equality if and only if $I(A,\a;B,\b)=I(A;B)$.
\end{theorem}
\begin{proof}
    Rearrange Lemma \ref{Lemma_ABSum} to get
    \begin{equation*}
        I(\a;\b)=I(A;\a)+I(B;\b)-I(A,B;\a,\b)+[I(A,\a;B,\b)-I(A;B)],
    \end{equation*}
    The Lemma now follows readily from the preceding expression:
    We obtain equality in the Lemma if (and only if) the
    term in brackets is zero.
    Otherwise, the bracketed term is nonnegative, since
    \begin{equation*}
    \begin{split}
        I(A,\a;&{} B,\b)- I(A;B) \\
        &{}=H(\a|A)+H(\b|B)-H(\a,\b|A,B) \\
        &{}=H(\a|A)-H(\a|A,B)+H(\b|B)-H(\b|A,B,\a)\\
        &{}\geq 0,
    \end{split}
    \end{equation*}
    where the inequality is due to the fact that conditioning reduces
    entropy.
\end{proof}

\begin{theorem}[Theorem \ref{Thm: Gelfand Pinsker lemma}]\label{Lemma_AInd}
    If $U_i=(\g,A^{i-1})$, then
    \begin{equation*}
        I(A^n;\g) = \sum_{i=1}^n I(A_i;U_i)- \sum_{i=2}^n I(A_i;A^{i-1})
    \end{equation*}
\end{theorem}
\begin{proof}
    In Lemma \ref{Lemma_ANoInd}, put $A=A_i$, $\a=A^{i-1}$. Note that $U_1=\g$.
    Hence, substituting and summing from $2$ to $n$ yields
        \begin{eqnarray*}
            \sum_{i=2}^n I(A_i;A^{i-1})&=&\sum_{i=2}^n
            I(A_i;U_i)-I(A^n;\g)+I(A_1;\g) \\
            &=& \sum_{i=2}^n I(A_i;U_i)-I(A^n;\g) +I(A_1;U_1)\\
            &=& \sum_{i=1}^n I(A_i;U_i)-I(A^n;\g).
        \end{eqnarray*}
\end{proof}

\section{Simplification of convex hulls}\label{appendix: Convex hulls} In this section we argue
geometrically that the expressions for convex hulls of the inner
bound regions simplify to just one term in both the binary and
Gaussian cases. To discuss both cases simultaneously, let us
represent the surface of either inner bound by a positive valued
function $f: \calD \rightarrow \mathbb{R}_+$. Here, $\calD$ is a
square region
\begin{equation*}
\calD = \{r=(x,y)\in \mathbb{R}^2: 0\leq x \leq M, \; 0\leq y \leq
M \},
\end{equation*}
and $M$ is a positive constant. In the binary case, $f(r)=g(r)$,
and $D = [0,1]\times [0,1]$; in the Gaussian case, $f(r)=G(r)$,
and $D = [0,\infty)\times [0,\infty)$. Some important properties
shared by both cases are that for all $r=(x,y)\in\calD$,
\begin{equation*}
\begin{split}
&{} f(x,y) \geq 0,\;\; f(0,y)=f(x,0) = 0, \\
&{} f_x(r),f_y(r) > 0,\;\; f_{xx}(r),f_{yy}(r) < 0, \\
\end{split}
\end{equation*}
where the subscripts denote partial derivatives.

Denote the convex hull of $f(r)$ by $c(r)$. Generically, the
boundary of the convex hull is
\begin{equation*}
c(r) = \max \theta f(r_1)+\tb f(r_2),
\end{equation*}
where the maximum is over all triples $(\theta,r_1,r_2)$ such that
$r=\theta r_1+\tb r_2$, $\theta \in [0,1]$, and $r_1,r_2\in\calD$.
However, as argued next, for the cases under study this simplifies
to
\begin{equation*}
c(r) = \max \;\theta f(r'),
\end{equation*}
where $r=\theta r'$.

The convex hull of a surface can be characterized in terms of its
tangent planes. Given any point $r'=(x,y)\in\calD$, if its tangent
plane lies entirely above the surface, then $(r',f(r'))$ is on the
convex hull. If the tangent plane cuts \emph{through} the surface
at one or more other points, then $(r,f(r))$ is not on the convex
hull. If the tangent plane intersects the surface at exactly two
points, then both points are on the convex hull.

The tangent plane at an arbitrary point $r'=(x',y')\in\calD$ is
the set of points satisfying
\begin{equation*}
z(x,y) = f_x(x-x')+f_y(y-y')+z',
\end{equation*}
where the partial derivatives are evaluated at $r'$, i.e. $f_x =
f_x(r')$, $f_y = f_y(r')$, and $z'=f(r')$. The tangent plane
intersects the $z=0$ plane in a line. Setting $z(r)=0$ and solving
\begin{eqnarray*}
 y &=& m x + b, \;\; \text{where}  \\
 m &=& -(f_x/f_y) \\
 b &=& 1/f_y[ x'f_x +y'f_y - z'].
\end{eqnarray*}
Since $f_x,f_y>0$, the slope $m=-(f_x/f_y)$ is negative. This line
intersects the positive orthant whenever the intercept $b\geq0$,
in which case the tangent plane cuts through the surface, since
$f\geq0$. Thus, the only points on the original surface $f(x,y)$
that can be on the convex hull are those for which $b \leq 0$.

Next consider any path through $\D$ along a line segment $y=\alpha
x$, $\alpha>0$, starting from one of the `outer edges' of $\D$,
where $x=M$ or $y=M$, and consider what happens to the tangent
plane's line of intersection $\ell$  with the $z=0$ plane as we
move in along the path toward the origin $(0,0)$. Initially, the
tangent planes lie entirely above the surface, and the intercept
of $\ell$ is negative, $b<0$. This intercept increases along the
path until $b=0$, at which point $\ell$ intersects $(0,0)$. Here,
the tangent plane contains a line segment attached on one end to
the point of tangency, and at the other end to the point
$(r,f(r))=(0,0,0)$; everywhere else, the tangent plane is above
the surface. Continuing toward the origin, all other points along
the path have tangent planes such that $\ell$ has a positive
intercept $b>0$, hence these points are excluded from the convex
hull.

These considerations imply that the convex hull $c(r)$ is composed
entirely of two kinds of points. First, points which coincide with
the original surface, $c(r)=\theta f(r)$, with $\theta=1$. These
points occur at values of $r=(x,y)$ `up and to the right' of
$(0,0)$. Second, points along line segments connecting surface
points `up and to the right' $(r',f(r'))$ with the point
$(r,f(r))=(0,0,0)$, that is $c(r) = \theta f(r')+\tb f(0,0) =
\theta f(r')$, where $r=\theta r'$ and $\theta\in[0,1]$. Hence,
for all $r\in\D$, $c(r)$ has the desired form.

An example of another function that behaves in the same way just
described is $f(x,y) = (1-(1-x)^2)(1-(1-y)^2)$, with $\calD =
[0,1]\times [0,1]$.

\section{Properties of Gaussian mutual information}\label{appendix: Gaussian properties} Our analysis of
the Gaussian pattern recognition problem relies on well-known
results, stated below without proof.
\begin{lemma}\label{lemma: Gaussian info}
The mutual information between two Gaussian random vectors
$\boldX$ and $\boldY$ depends only on the matrices of correlation
coefficients. Specifically,
\begin{eqnarray*}
I(\boldX ; \boldY ) &=& \frac{1}{2} \log \left( \det \boldC_{x,x}
\right) - \frac{1}{2} \log \left( \det \boldC_{x,x|y} \right)  ,
\end{eqnarray*}
where
\begin{eqnarray*}
\boldC_{x,x|y} & = & \boldC_{x,x} - \boldC_{x,y} \boldC_{y,y}^{-1}
\boldC_{y,x} .
\end{eqnarray*}
\end{lemma}
In the most well known special case of $Y=X+W$, where $X$ and $W$
are independent Gaussian random variables with variances $P$ and
$N$, respectively, yields
\begin{equation*}
I(X;Y)  =  \frac{1}{2} \log\left(1+ \frac{P}{N} \right)
 =  - \frac{1}{2} \log\left(1 - \rho_{x,y}^2\right) ,
\end{equation*}
where the correlation coefficient $\rho_{x,y} = \sqrt{P/(P+N)}$.

\begin{lemma}\label{lemma: gauss-markov} If $X,Y$ and $Z$ are zero mean Gaussian random vectors that form a
Markov chain $X-Y-Z$, then
\begin{equation*}
\boldC_{x,z}  =  \boldC_{x,y} \boldC_{y,y}^{-1} \boldC_{y,z}.
\end{equation*}
\end{lemma}
Note that for dimension one, $X \rightarrow Y \rightarrow Z$
implies $\rho_{x,z} = \rho_{x,y} \rho_{y,z}$.

\begin{lemma}\label{lemma: length four correlations}
Let $X, Y$, $U$, and $V$ be jointly Gaussian random variables such
that $U -X -Y$ and $X  - Y - V$ are Markov chains. Then the matrix
of correlation coefficients $C_{xy,uv}$ decomposes as
\begin{equation*}
C_{xy,uv} = \left[
\begin{array}{cc}
1 \ &  \ \rxy \\
\rxy \ &  \ 1
\end{array}
\right] \ \left[
\begin{array}{cc}
\rxu  \ &  \ 0 \\
0  \ &  \ \ryv
\end{array}
\right].
\end{equation*}
\end{lemma}
This lemma follows immediately by using Lemma \ref{lemma:
gauss-markov} to obtain the substitutions $\boldC_{x,v} =
\boldC_{x,y} \boldC_{y,y}^{-1} \boldC_{y,v}=\rxy\ryv$ and
$\boldC_{u,y} = \boldC_{u,x} \boldC_{x,x}^{-1}
\boldC_{x,y}=\rux\rxy$.

\bibliography{a_BrandonsBibtexFile,CoverThomasBibTex}

\end{document}